\documentclass[prb,aps,twocolumn,showpacs,amsmath,floatfix,superscriptaddress]{revtex4}
\usepackage{graphicx,epsf,natbib,amsmath,latexsym,amssymb}
\def\be{\begin{equation}}
\def\ee{\end{equation}}
\def\bea{\begin{eqnarray}}
\def\eea{\end{eqnarray}}

\begin{document}
\title{Nonequilibrium transport of helical Luttinger liquids through a quantum dot}

\author{Sung-Po Chao}
\affiliation{Physics Division, National Center for Theoretical Science, Hsinchu, 30013, Taiwan, R.O.C.}
\affiliation{Physics Department, National Tsing Hua University, Hsinchu, 30013, Taiwan, R.O.C.}
\author{Salman A. Silotri}
\affiliation{Electrophysics Department, National Chiao-Tung University, Hsinchu, 30010, Taiwan, R.O.C.}
\author{Chung-Hou Chung}
\affiliation{Physics Division, National Center for Theoretical Science, Hsinchu, 30013, Taiwan, R.O.C.}
\affiliation{Electrophysics Department, National Chiao-Tung University, Hsinchu, 30010, Taiwan, R.O.C.}
\date{\today}

\begin{abstract}
We study a steady state non-equilibrium transport between two interacting helical edge states of a two dimensional topological insulator, described by helical Luttinger liquids, through a quantum dot. For non-interacting dot the current is obtained analytically by including the self-energy correction to the dot Green's function. For interacting dot we use equation of motion method to study the influence of weak on-site Coulomb interaction on the transport. We find the 
metal-to-insulator quantum phase transition for attractive \emph{or} repulsive interactions in the leads when the magnitude of the interaction strength characterized by a charge sector Luttinger parameter $K$ goes beyond a critical value. The critical Luttinger parameter $K_{cr}$ depends on the hoping strength between dot and the leads as well as the energy level of the dot with respect to the Fermi levels of the leads, ranging from weak interaction regime for dot level off resonance to strong interaction regime for dot in resonance with the equilibrium Fermi level. Nearby the transition various singular behaviors of current noise, dot density of state, and the decoherence rate (inverse of lifetime) of the dot are briefly discussed.   
\end{abstract}
\maketitle

\section{Introduction}
 The topological properties of quantum matter have attracted a great deal of attentions in 
condensed matter systems since the discovery and comprehensive study of quantum Hall effect.
In systems with time-reversal symmetry and strong spin-orbit interactions the quantum spin 
Hall insulator (QSHI) has been theoretically proposed\cite{Kane0, Zhang, Zhang2} and soon afterwards, experimentally verified  
the existence of the topologically non-trivial edge states, the hallmark of QSHI, in HgTe/CdTe quantum well structures\cite{Konig,Molenkamp}. The QSHI is a time reversal invariant two dimensional electronic phase which has a bulk energy gap generated by spin-orbit interaction. The topological order\cite{Mele} of the this state, similar to the case of integer quantum Hall effect, requires the presence of gapless edge states. The propagation direction at one edge is opposite for opposite
spins, and thus the edge states are usually named as the helical liquids. This one dimensional edge state is protected from elastic back scattering through time reversal symmetry as the backscattering requires spin flips. Recent experiments have verified the perfectly transmitted\cite{Konig} Landauer conductance $2e^2/h$ as well as the spin orientations of the edge state transport through junction device\cite{Molenkamp,David} and theoretical proposals for detecting spin orientations through spin polarized scanning tunneling microscope\cite{stm}. 

In the presence of electron-electron interactions, these one dimensional helical edge states form "helical Luttinger liquids"\cite{HLL}, a new type of Luttinger liquids where the spins of electrons are tied to the directions of their momenta. This helical nature of QSHI edge states leads to various different transport properties from the ordinary or chiral Luttinger liquids realized as edge states of integer quantum Hall system. 

The transport properties of the helical Luttinger liquids have been discussed in the setup of a quantum point contact
\cite{Hou,Anders,Teo,Liu} between the edges. 
Alternatively, transport of a quantum dot\cite{Law, Seng} or antidot\cite{Posske,Dolcetto} coupled to the helical 
edge states offer a simple way for detection of the 
helical Luttinger liquids. More interestingly, 
when quantum dot is in the Kondo regime, the setup could be used to probe the transition (or crossover) between one-channel to two-channel Kondo physics\cite{Law} by changing the interaction strength or impurities concentrations of the edge state. 
The idea of using repulsive interaction of the electrons in the leads to suppress Kondo couplings between the two leads was suggested earlier by Fabrizio and Gogolin\cite{Fabrizio} for the case of two Luttinger liquid leads coupled to a quantum dot. There they show that the two-channel Kondo can be reached with Luttinger parameter $K<1/2$. In the work by Law et al.\cite{Law} they show this two-channel fixed point can be reached nearby equilibrium for weaker repulsive interaction (with Luttinger parameter $K<1$), in the context a quantum dot coupled to two helical Luttinger liquid leads realized as edge states of the QSHI. Higher order of renormalization group analysis\cite{Chung} shows there could be a quantum phase transition between one-channel and two-channel Kondo in the same setup for Luttinger parameter $1/2<K<1$.  

\begin{figure}
\includegraphics[width=.7\columnwidth]{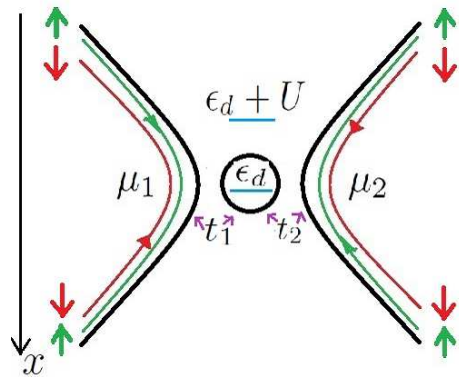}
\caption{Nonequilibrium steady state transport between two edge states of quantum spin Hall system through a quantum dot. Dot level $\epsilon_d$ is controlled by the gate voltage applied on the quantum dot. The coupling strengths between dot and the edge states leads $t_1$ and $t_2$ can also be varied in the experiment. $U$ represents the on dot Coulomb interaction. Small solid colored arrows indicates spin orientations.}
 \label{f0}
\end{figure}

 In this paper, we address a different aspect (in the resonant tunneling limit far from the Kondo regime) of the quantum dot setup in Ref.~\onlinecite{Law}: we study the nonequilibrium steady state transport problem with a non-interacting or weakly interacting quantum dot connected with two helical Luttinger liquid edges of the QSHI as shown in the Fig. \ref{f0}. 
The problem of non-interacting quantum dot connected to leads of chiral Luttinger liquids was studied by Chamon and Wen\cite{Chamon} and later generalized to multi-level within the quantum dot by Furusaki\cite{Furusaki} using master equation approach at temperature higher than the tunneling strength. The nonequilibrium transport of a non-interacting quantum dot coupled to one side of the helical edge state and a normal Fermi liquid lead was studied by Seng and Ng in Ref.~\onlinecite{Seng}. For a quantum dot connected with two edges kept at different chemical potentials we map this problem into spinful Luttinger liquids following Hou et al. in Ref.~\onlinecite{Hou}. While helicity makes spin a redundant quantum number on a single edge\cite{Seng}, it is important to include it when two edges are connected via a quantum dot\cite{Anders}. In this mapping the charge sector Luttinger parameter $K_c$ is connected with the spin sector of Luttinger parameter $K_s$ by $K_c=1/K_s$, a unique property owing to the helical nature of interacting edge state of QSHI.
Charge and spin susceptibility measurement\cite{Giamarchi} can be used to probe these two quantities independently and confirm their connections. 

\begin{figure*}
\begin{center}$
\begin{array}{cc}
\includegraphics[width=1\columnwidth]{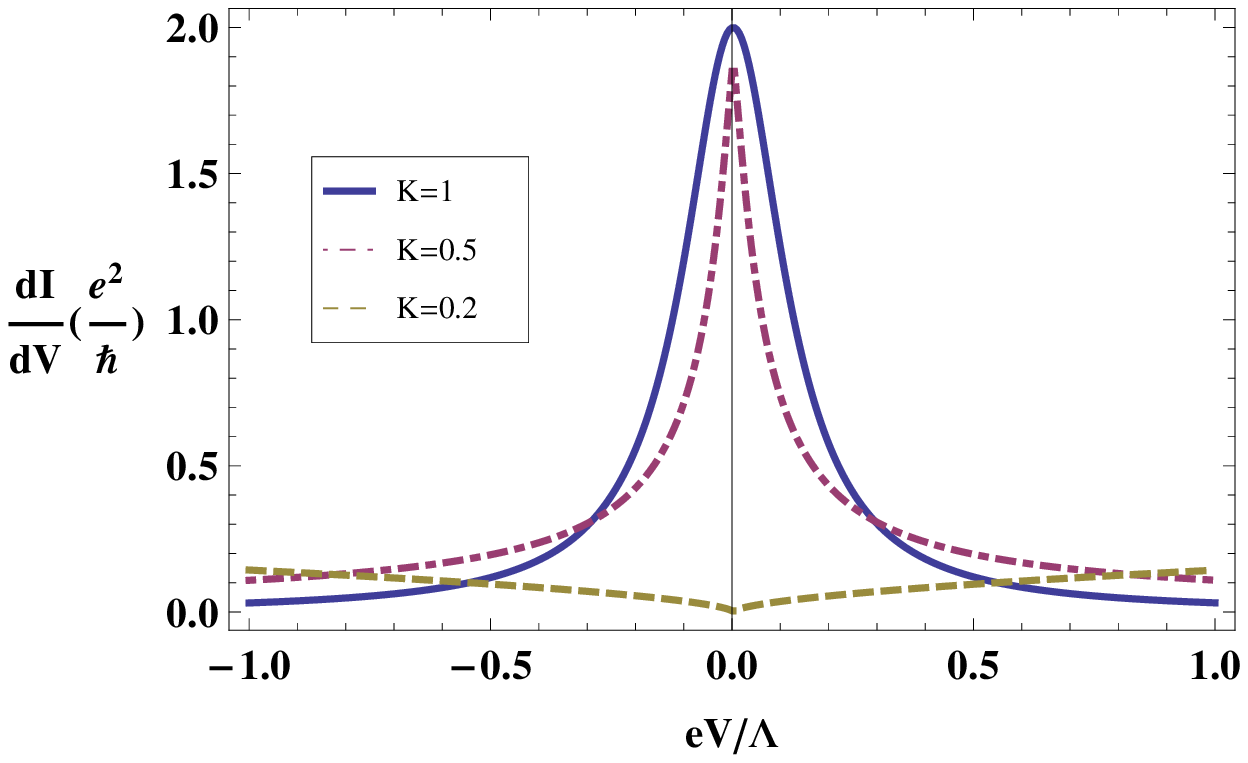}&
\includegraphics[width=1\columnwidth]{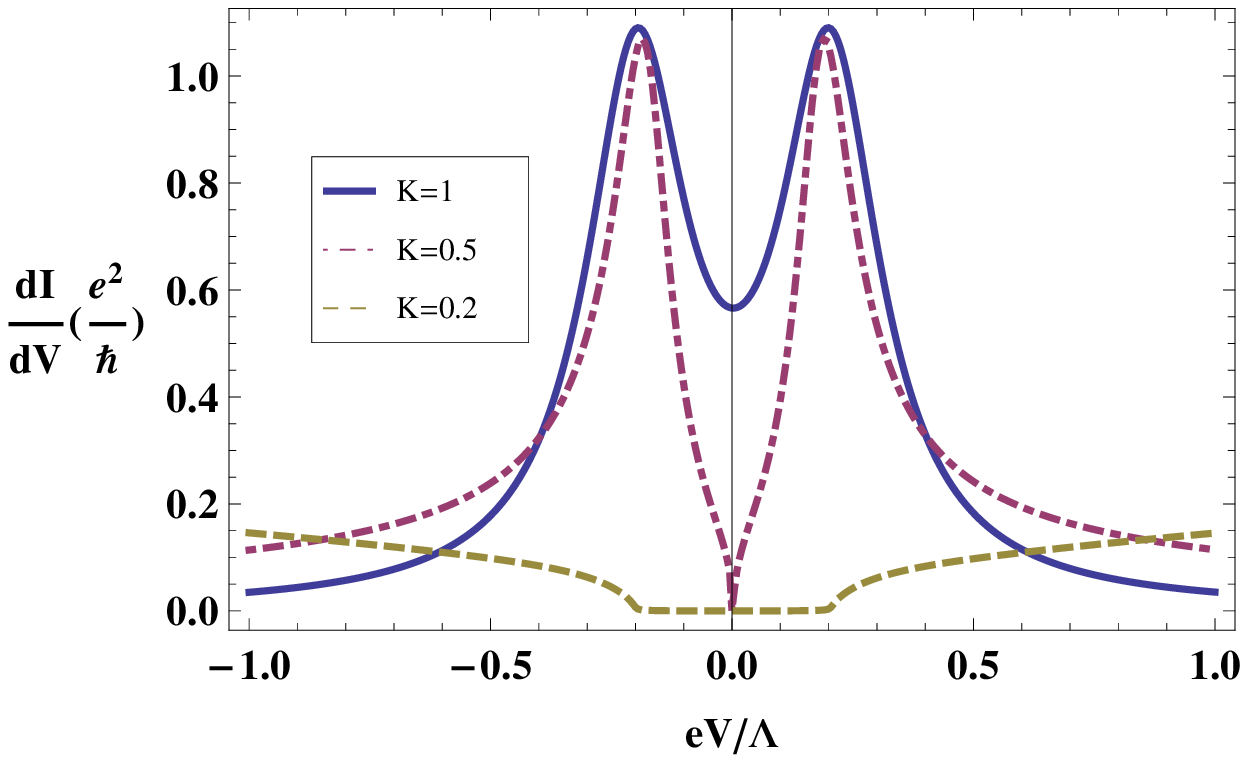}
\end{array}
$\end{center}
\caption{Differential conductance as a function of voltage obtained numerically for $K=1$ (blue line), $K=0.5$ (purple dot dashed line), $K=0.2$ (brown dashed line), and $t/\Lambda=0.1$. Left: $\epsilon_d/\Lambda=0$ and Right: $\epsilon_d/\Lambda=-0.1$. Note that the insulating behavior at $V=0$ occurs at larger $K$ value (for repulsive interaction, the $K=0.5$ is already in the insulating phase at zero voltage) compared with $\epsilon_d=0$ case.}
 \label{f1}
\end{figure*}

\begin{figure*}
\begin{center}$
\begin{array}{cc}
\includegraphics[width=1\columnwidth, clip]{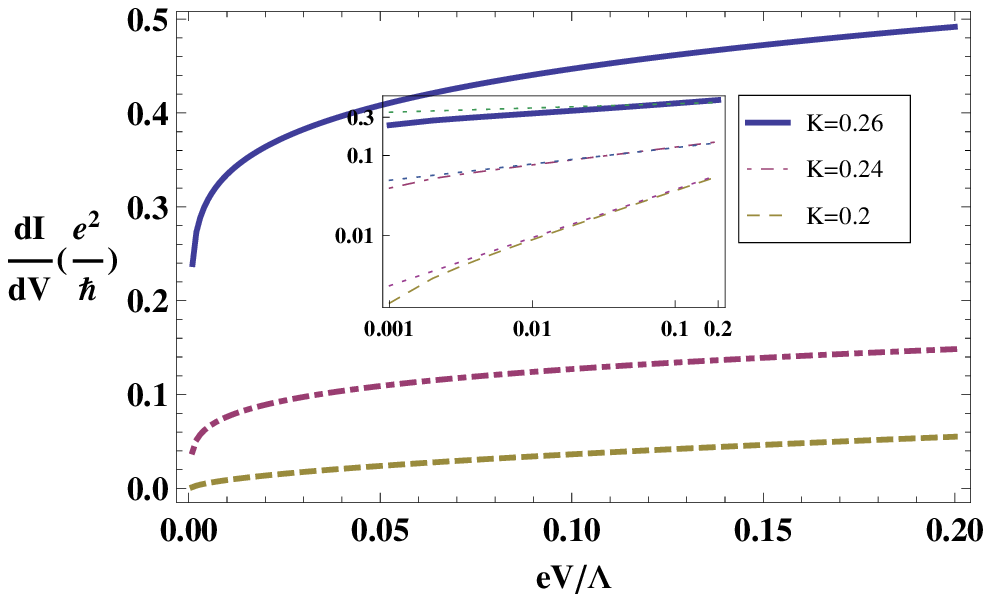}&
\includegraphics[width=1\columnwidth, clip]{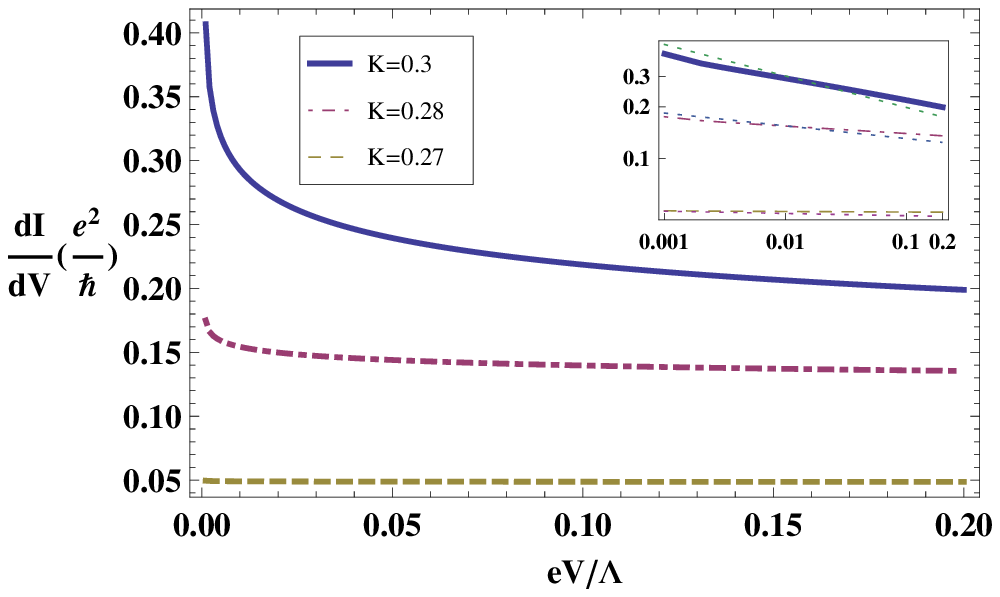}
\end{array}
$\end{center}
\caption{$dI/dV$ v.s. $V$ for various interaction strength with $t/\Lambda=0.1$, $\epsilon_d=0$ and $\mu_1=-\mu_2=\frac{eV}{2}$. Inset shows double logarithmic plot, illustrating the power law behavior for conductivity ($dI/dV\propto V^{\frac{1}{2}\left(K+\frac{1}{K}\right)-2}$) slightly away from the resonance value $eV\simeq \epsilon_d$. The straight dotted lines are generated for guidance by functions proportional to $V^{\frac{1}{2}\left(K+\frac{1}{K}\right)-2}$ for different Luttinger parameter $K$. Left: Insulating phase for $K\le K_{cr}\sim 0.267$: $K=0.26$ (blue line), $K=0.24$ (purple dot dashed line), $K=0.2$ (brown dashed line). Right: Conducting phase for $K\ge K_{cr}$: $K=0.3$ (blue line), $K=0.28$ (purple dot dashed line), $K=0.27$ (brown dashed line).}
 \label{f2}
\end{figure*}

For non-interacting dot the dot Green's function is obtained exactly through inclusion of all order of perturbation and the charge current is expressed analytically through Keldysh perturbation\cite{Jauho, Haug}. The differential conductance is obtained by numerical derivative on the current-voltage relation. We find at zero temperature the zero voltage conductance width and height decreases with increasing repulsive or attractive interactions within the edges, similar to that for other type of Luttinger liquids leads\cite{Kane,Kane1}. The equilibrium conductance reaches zero for strongly interacting edge states (with Luttinger parameter $K<0.26$ for repulsive interaction) for the dot level on resonance with the equilibrium Fermi surface of the edges. We compute other physical measurable quantities such as noise\cite{Schmidt,Simon,Yuwen} and lifetime of the dot electron and both of them show similar transitions at the same interaction strength in
lowest order perturbation computation. For dot level away from the equilibrium Fermi level this metallic to insulating, or quantum phase transition (QPT), occurs at weaker interaction strengths as can be seen by comparing Fig.\ref{f1} and Fig.\ref{f2}. Note that for very strong repulsive interactions (when Luttinger parameter $K<1/4$) random two-particle back scatterings, albeit preserving time reversal symmetry, destablize the edge states\cite{Wu,Xu}. Magnetic impurities along the edge also destablize the edge states\cite{Maciejko} for $K<1/4$. This hinders the possibility of observing the QPT in experiment if the dot level is at resonance with the equilibrium Fermi sea but we shall be able to observe this QPT by tunning the dot level off resonance. \emph{Dot level, controlled by the gate voltage upon it, serves as another tunable parameter in additions to interaction strengths of the edge states to drive this QPT.}

For weakly interacting dot we use equation of motion approach to compute perturbatively the influence of Coulomb repulsion for double occupancy on the quantum dot. For dot level at resonance we again find the metal-to-insulator transition at the same critical interaction strength $K_{cr}\sim 0.26$ as that for a non-interacting dot. 

This article is organized as follows. In Sec. \ref{st2}, we setup the model Hamiltonian and apply bosonization\cite{Giamarchi,gogolin} to solve for the edge state Hamiltonian before coupling it to the quantum dot. In the presence of a quantum dot, we use Keldysh perturbation theory to compute the charge current through the quantum dot. The calculations for non-interacting dot on the current, noise through the quantum dot, and the lifetime of the dot electron are summarized in Sec. \ref{st3}. In Sec. \ref{st4} we compute the current transport through weakly interacting dot. The last section is devoted to the conclusion. Derivations of various correlators of the edge states are shown in the Appendix. \ref{a1}.

\section{Model Hamiltonian}\label{st2}
Following Hou et al. in Ref.~\onlinecite{Hou} we model the two edge states of the QSHI kept at chemical potentials $\mu_1$ and $\mu_2$ connected via a quantum dot as
\begin{eqnarray}\label{eq1}
&&H=H_{edges}+H_{dot}+H_{int}=H_0+H_{int}\\\nonumber
&&H_{edges}=\\\nonumber &&\int dx \Big\{\sum_{\sigma}\Big[iv_F (\psi_{L\sigma}^{\dagger}(x)\partial \psi_{L\sigma}(x)-\psi_{R\sigma}^{\dagger}(x)\partial \psi_{R\sigma}(x))\\\nonumber
&&+u_2\psi_{L\sigma}^{\dagger}(x)\psi_{L\sigma}(x)\psi_{R-\sigma}^{\dagger}(x)\psi_{R-\sigma}(x)+\frac{u_4}{2}\big(\psi_{L\sigma}^{\dagger}(x)\psi_{L\sigma}(x)\\\nonumber&&\times\psi_{L\sigma}^{\dagger}(x)\psi_{L\sigma}(x)+\psi_{R\sigma}^{\dagger}(x)\psi_{R\sigma}(x)\psi_{R\sigma}^{\dagger}(x)\psi_{R\sigma}(x)\big)\Big]
\\\nonumber
&&-\mu_1\big(\psi_{R\uparrow}^{\dagger}(x)\psi_{R\uparrow}(x)+\psi_{L\downarrow}^{\dagger}(x)\psi_{L\downarrow}(x)\big)-\mu_2\big(\psi_{L\uparrow}^{\dagger}(x)\psi_{L\uparrow}(x)\\\nonumber
&&+\psi_{R\downarrow}^{\dagger}(x)\psi_{R\downarrow}(x)\big)\Big\}
\end{eqnarray}
\begin{eqnarray*}
&&H_{dot}=\sum_{\sigma}\epsilon_d d_{\sigma}^{\dagger}d_\sigma+U d_{\uparrow}^{\dagger}d_{\uparrow}d_{\downarrow}^{\dagger}d_{\downarrow}\\\nonumber
&&H_{int}=[t_1(\psi_{R\uparrow}^{\dagger}(0)d_{\uparrow}+\psi_{L\downarrow}^{\dagger}(0)d_{\downarrow}+h.c.)\\\nonumber&&+t_2(\psi_{L\uparrow}^{\dagger}(0)d_{\uparrow}+\psi_{R\downarrow}^{\dagger}(0)d_{\downarrow}+h.c.)]
\end{eqnarray*}
Here $u_2$, $u_4$ are the interaction constants modeling the short range interaction within the edge. $\epsilon_d$ is the dot energy level and $U$ is the on dot Coulomb interaction. $t_1$ and $t_2$ are the coupling strengths between the edges and the dot. Using the spinful bosonization by writing the fermion fields of the edge states as
\begin{eqnarray*}
\psi_{L\sigma}(x)=\frac{1}{\sqrt{2\pi a_0}}\eta_{\sigma}e^{-i\sqrt{4\pi}\phi_{L\sigma}(x)},\\
\psi_{R\sigma}(x)=\frac{1}{\sqrt{2\pi a_0}}\eta_{\sigma}e^{-i\sqrt{4\pi}\phi_{R\sigma}(x)},
\end{eqnarray*}  
with $\eta_{\sigma}$ as the Klein factor chosen to satisfy the fermion anti-commutation rule and $a_0$ as the lattice cutoff for the linear spectrum. Define 
the bosonic fields 
\begin{eqnarray*}
\Phi_{\sigma}/\Theta_{\sigma}=\phi_{L\sigma}\pm\phi_{R\sigma}
\end{eqnarray*}
and denote their charge and spin sectors as 
\begin{eqnarray*}
\Phi_c=\frac{1}{\sqrt{2}}(\Phi_{\uparrow}+\Phi_{\downarrow}) , \Phi_s=\frac{1}{\sqrt{2}}(\Phi_{\uparrow}-\Phi_{\downarrow})
\end{eqnarray*}
and the similar expressions for $\Theta_c$ and $\Theta_s$. We choose a time dependent gauge transformation to move the chemical potentials in $H_0$
to $H_{int}$ by writing
\begin{eqnarray*}
\psi_{R\uparrow}/\psi_{L\downarrow}\rightarrow e^{i\mu_1 t}\psi_{R\uparrow}/\psi_{L\downarrow} \\
 \psi_{L\uparrow}/\psi_{R\downarrow}\rightarrow e^{i\mu_2 t}\psi_{L\uparrow}/\psi_{R\downarrow} 
\end{eqnarray*}
With these transformations we rewrite Eq.(\ref{eq1}) as
\begin{eqnarray}\nonumber
H_0&=&\sum_{\alpha=c,s}\frac{v_\alpha}{2}\int_{-\infty}^{\infty} dx :[K_\alpha(\partial_x \Theta_\alpha)^2+\frac{1}{K_\alpha}(\partial_x \Phi_\alpha)^2]:\\
&+&\sum_{\sigma}\epsilon_d d_{\sigma}^{\dagger}d_\sigma+U d_{\uparrow}^{\dagger}d_{\uparrow}d_{\downarrow}^{\dagger}d_{\downarrow}\\\nonumber
H_{int}&=&\sum_{\sigma}(t_{R\sigma}e^{-i\mu_{R\sigma} t-i\sqrt{4\pi}\phi_{R\sigma}(0)}\eta_{R\sigma}^{\dagger}d_{\sigma}\\\nonumber&+&t_{L\sigma}e^{-i\mu_{L\sigma} t-i\sqrt{4\pi}\phi_{L\sigma}(0)}\eta_{L\sigma}^{\dagger}d_{\sigma}+h.c.)
\end{eqnarray}
with $K_c=1/K_s=K=\sqrt{\frac{1+\frac{u_4}{2\pi v_F}-\frac{u_2}{2\pi v_F}}{1+\frac{u_4}{2\pi v_F}+\frac{u_2}{2\pi v_F}}}$, $v_c=v_s=v=v_F\sqrt{(1+\frac{u_4}{2\pi v_F})^2-(\frac{u_2}{2\pi v_F})^2}$, $t_{R\uparrow}=t_{L\downarrow}=t_1/\sqrt{2\pi a_0}$, $t_{L\uparrow}=t_{R\downarrow}=t_2/\sqrt{2\pi a_0}$, $\mu_{R\uparrow}=\mu_{L\downarrow}=\mu_1$, and $\mu_{L\uparrow}=\mu_{R\downarrow}=\mu_2$. 
For repulsive interaction in the edge states $u_2>0$ and $u_4>0$ which leads to Luttinger parameter $K<1$. For attractive interaction we have Luttinger parameter $K>1$. The particle current at time $t$, or $I_1(t)/e=-I_2(t)/e$ with $e$ as electric charge, is obtained by the Heisenberg equation:
\begin{eqnarray*}
&&I_1(t)/e=d \langle \psi_{R\uparrow}^{\dagger}(0)\psi_{R\uparrow}(0)+\psi_{L\downarrow}^{\dagger}(0)\psi_{L\downarrow}(0)\rangle/dt\\
&&=-i\langle[\psi_{R\uparrow}^{\dagger}(0)\psi_{R\uparrow}(0)+\psi_{L\downarrow}^{\dagger}(0)\psi_{L\downarrow}(0), H]\rangle\\&&=2t_1\Im[\langle e^{-i\mu_{R\uparrow} t}\psi^{\dagger}_{R\uparrow}(0)d_\uparrow+e^{-i\mu_{L\downarrow} t}\psi^{\dagger}_{L\downarrow}(0)d_\downarrow\rangle]\\&&=
2\Im[\langle t_{R\uparrow}e^{-i\sqrt{4\pi}\phi_{R\uparrow}(0,t)}\eta_{R\uparrow}^{\dagger}d_{\uparrow}+t_{L\downarrow}e^{ -i\sqrt{4\pi}\phi_{L\downarrow}(0,t)}\eta_{L\downarrow}^{\dagger}d_{\downarrow}\rangle]
\end{eqnarray*}
Here $\phi_{R/L\sigma}(0,t)\equiv\phi_{R/L\sigma}(0)+\mu_{R/L\sigma}t/\sqrt{4\pi}$. 
 By defining the Keldysh contour ordered Green's function $G_{\sigma,R/L\sigma}(t,t')=-i\langle T_c\{d_\sigma(t)\psi^{\dagger}_{R/L\sigma}(t')\}\rangle$ we express the particle current as
\begin{eqnarray}
 I(t)/e&=&(I_1(t)-I_2(t))/2e\\\nonumber&=&\Re[\sum_{j=R,L;\sigma}(-1)^{j\sigma}t_{j\sigma}e^{-i\mu_{j\sigma}t}G_{\sigma,j\sigma}^{<}(t,t)].
\end{eqnarray}
The sign $(-1)^{j\sigma}$ is chosen as: $(-1)^{L\uparrow}=(-1)^{R\downarrow}=-1$ and $(-1)^{R\uparrow}=(-1)^{L\downarrow}=1$.
 This lesser mixed Green's function $G_{\sigma,j\sigma}^{<}(t,t)$ is obtained by analytic continuation of contour ordered Green's function $G_{\sigma,R/L\sigma}(\tau,\tau')$ in imaginary time $\tau$ and $\tau'$ with
its expression given by perturbation as
\begin{eqnarray}\nonumber
&&G_{\sigma,R/L\sigma}(\tau,\tau')=\sum_{l=0}^{\infty}\frac{(-i)^{l+1}}{l!}\int_c d\tau_1\ldots\int_c d\tau_l\langle T_c\{d_{\sigma}(\tau)\\\label{dg}&&H_{int}(\tau_1) \ldots
H_{int}(\tau_l)\psi^{\dagger}_{R/L\sigma}(\tau')\}\rangle
\end{eqnarray}
In applying the Wick theorem in the Eq.(\ref{dg}) we should also include all possible four fermions interactions term ($u_2$ and $u_4$ term in the edge states Hamiltonian) between any two fermions operators. We use the spinful bosonization as a way to sum up all orders of perturbations in the four fermions interactions in Keldysh form. The edge state correlators evaluated this way is thus fully dressed in our treatment and we do not specify this aspect in the expression of Eq.(\ref{dg}). 

 In the following two sections we solve this mixed Green's function perturbatively in the case of noninteracting dot and use equation of motion approach to
study the case of weak on dot interaction $U$ at zero temperature.  

\section{Noninteracting dot}\label{st3}
\subsection{Charge current}
For noninteracting quantum dot ($U=0$) the $H_0$ is quadratic in dot electron operator $d_\sigma$ and
the lowest nonzero perturbation gives
\begin{eqnarray*}
&&G_{\sigma,R/L\sigma}(\tau,\tau')=-\int_c d\tau_1\langle  T_c\{d_{\sigma}(\tau)H_{int}(\tau_1)\psi^{\dagger}_{R/L\sigma}(\tau')\}\rangle
\end{eqnarray*}
which involves bare dot Green's function $i G_{d_\sigma}^{(0)}(\tau,\tau_1)=\langle T_c\{d_{\sigma}(\tau) d^{\dagger}_{\sigma}(\tau_1)\}\rangle$ and bare edge states Green's function $i G_{\psi_{R/L,\sigma}}(\tau,\tau_1)=\langle T_c\{\psi_{R/L\sigma}(\tau_1)\psi^{\dagger}_{R/L\sigma}(\tau')e^{-i\mu_{R/L\sigma}(\tau'-\tau_1)}\}\rangle$. The terminology "bare" here means the dot and leads are decoupled. The decoupled leads or edge states are described by fully interacting helical Luttinger liquids. From Eq.(\ref{dg}) we sum over all order of $H_{int}$ and the Fourier transformed full retarded dot Green's function is given by 
\begin{eqnarray*}
G_{d\sigma}^{R}(\omega)=G_{d\sigma}^{(0)R}(\omega)+G_{d\sigma}^{(0)R}(\omega)\Sigma_{\sigma}^R(\omega)G_{d\sigma}^{R}(\omega).
\end{eqnarray*}
 Similarly the full dot lesser Green's function is $G_{d\sigma}^{<}(\omega)=G_{d\sigma}^{R}(\omega)\Sigma_{\sigma}^{<}(\omega)G_{d\sigma}^{A}(\omega)$. Here the dot self energy is 
\begin{eqnarray}\label{se1}
\Sigma_{\sigma}(\omega)\equiv\sum_{j}|t_{j\sigma}|^2G_{\psi_j,\sigma}(\omega).
\end{eqnarray}
 The bare dot retarded Green's function is
$G_{d\sigma}^{(0)R}(\omega)=1/(w-\epsilon_d+i0^+)$.
The charge current is 
\begin{eqnarray}\nonumber
&&I(t)= e\Re[\sum_{j=R,L;\sigma}(-1)^{j\sigma}t_{j\sigma}e^{-i\mu_{\alpha\sigma}t}G_{\sigma,j\sigma}^{<}(t,t)]\\ \nonumber
&&=e\Re[\sum_{j,\sigma}(-1)^{j\sigma}|t_{j\sigma}|^2\int dt_1(G_{d_\sigma}^R(t,t_1)G_{\psi_{j,\sigma}}^{<}(t_1,t)\\
&&+G_{d_\sigma}^{<}(t,t_1)G_{\psi_{j,\sigma}}^{A}(t_1,t))]
\end{eqnarray}
For steady state $G_{\psi_{j,\sigma}}(t,t_1)=G_{\psi_{j,\sigma}}(t-t_1)$ and $G_{d_\sigma}(t,t_1)=G_{d_\sigma}(t-t_1)$. We rewrite steady state current $I(t)=\langle\hat{I}\rangle$ 
as
\begin{eqnarray}\nonumber
\langle\hat{I}\rangle&=&e\Re[\sum_{j,\sigma}(-1)^{j\sigma}|t_{j\sigma}|^2\int d\omega (G_{d_\sigma}^R(\omega)G_{\psi_{j,\sigma}}^{<}(\omega)\\ \label{curr}
&+&G_{d_\sigma}^{<}(\omega)G_{\psi_{j,\sigma}}^{A}(\omega))]
\end{eqnarray}

\begin{figure}
\includegraphics[width=1\columnwidth]{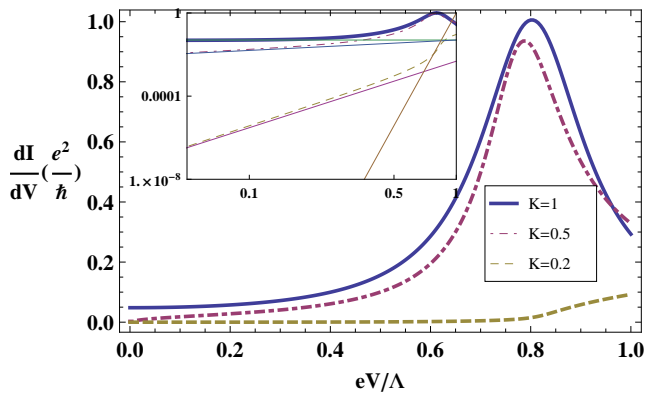}
\caption{Differential conductivity as a function of voltage obtained numerically for $K=1$ (blue line), $K=0.5$ (purple dot dashed line), and $K=0.2$ (brown dashed line). $\epsilon_d/\Lambda=-0.4$, $t/\Lambda=0.1$ and $\mu_1=-\mu_2=eV/2$. Inset shows the double logarithmic plot with the top three thin straight lines attached to the numerical data away from the resonance peak, generated for guidance by functions proportional to $V^{\left(K+\frac{1}{K}\right)-2}$ for the corresponding different Luttinger parameter $K$. The steepest (brown) thin straight line corresponds to two particle scattering process generated by function proportional to $V^{\left(\frac{4}{K}-2\right)}$ for $K=0.2$ case. A small region close to the resonance peak is described by this power law behavior.}
 \label{f1a0}
\end{figure}

The various $G_{\psi_{j,\sigma}}(\omega)$ are computed in the Appendix. \ref{a1} and we use them to obtain the full impurity Green's function. The current at zero temperature is evaluated numerically and the differential 
conductivity, obtained by taking numerical derivative on the current-voltage curves, for symmetrically coupled ($t_1=t_2$) and symmetrically driven voltage ($\mu_1=-\mu_2=eV/2$) is plotted in the left figure of Fig.\ref{f1} for $\epsilon_d=0$ case and the right figure of Fig.\ref{f1} for $\epsilon_d=-0.1\Lambda$. 
For asymmetrically driven voltage or $t_1\neq t_2$ the overall feature discussed below are similar but the symmetry between positive and negative voltage breaks.
Thus we concentrate on the case for $t_1=t_2$ and $\mu_1=-\mu_2$ in our analysis hereafter. For $\epsilon_d=0$, with $0$ as our equilibrium Fermi level, the scaling dimension obtained from zeroth order renormalization group (RG) analysis\cite{gogolin} in $H_{int}$ is $(K_c+K_s+1/K_c+1/K_s)/8$ which renders the renormalized 
coupling $t$ in equilibrium as
\begin{eqnarray}\label{eq8}
\frac{dt}{d\ln(\Lambda)}=\left(1-\frac{K_c+K_s+1/K_c+1/K_s}{8}\right)t
\end{eqnarray}
with $\Lambda=\hbar v/a_0$ as the UV cutoff for the linear spectrum of the edge states. For quantum spin Hall state $K_c=1/K_s=K$ and we have the critical value of $K_{cr}=2\pm\sqrt{3}$. That is, for  $0.267<K<3.733$ we have finite conductance at $V=0$ and insulating behavior for $K<0.26$ or $K>3.733$. To understand this critical behavior for dot level in resonance with the equilibrium Fermi surface we plot the differential conductance versus source drain voltage in Fig.\ref{f2} for Luttinger parameter $K$ chosen slightly below (left figure of Fig.\ref{f2}) and above (right figure of Fig.\ref{f2}) the critical Luttinger parameter $K_{cr}=2\pm\sqrt{3}$. The inset shows the double logarithm plots for both figures and in both cases shows $dI/dV\propto V^{\frac{1}{2}\left(K+\frac{1}{K}\right)-2}$ for voltage slightly off resonance. The drastic difference nearby $V\simeq0$ clearly illustrate the metal to insulator transition across the critical Luttinger 
parameter $K_{cr}\sim 0.26$ for dot level in resonance with the equilibrium Fermi level.

\begin{figure*}
\begin{center}$
\begin{array}{cc}
\includegraphics[width=1\columnwidth, clip]{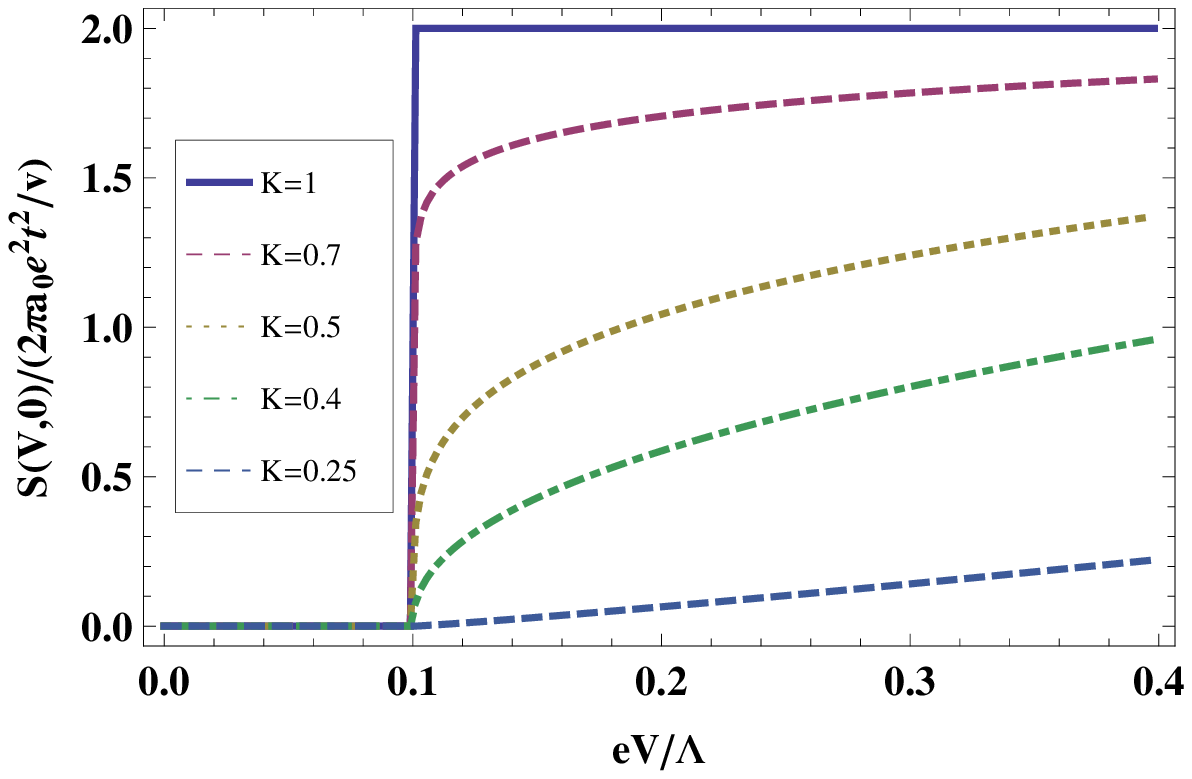}&
\includegraphics[width=1\columnwidth, clip]{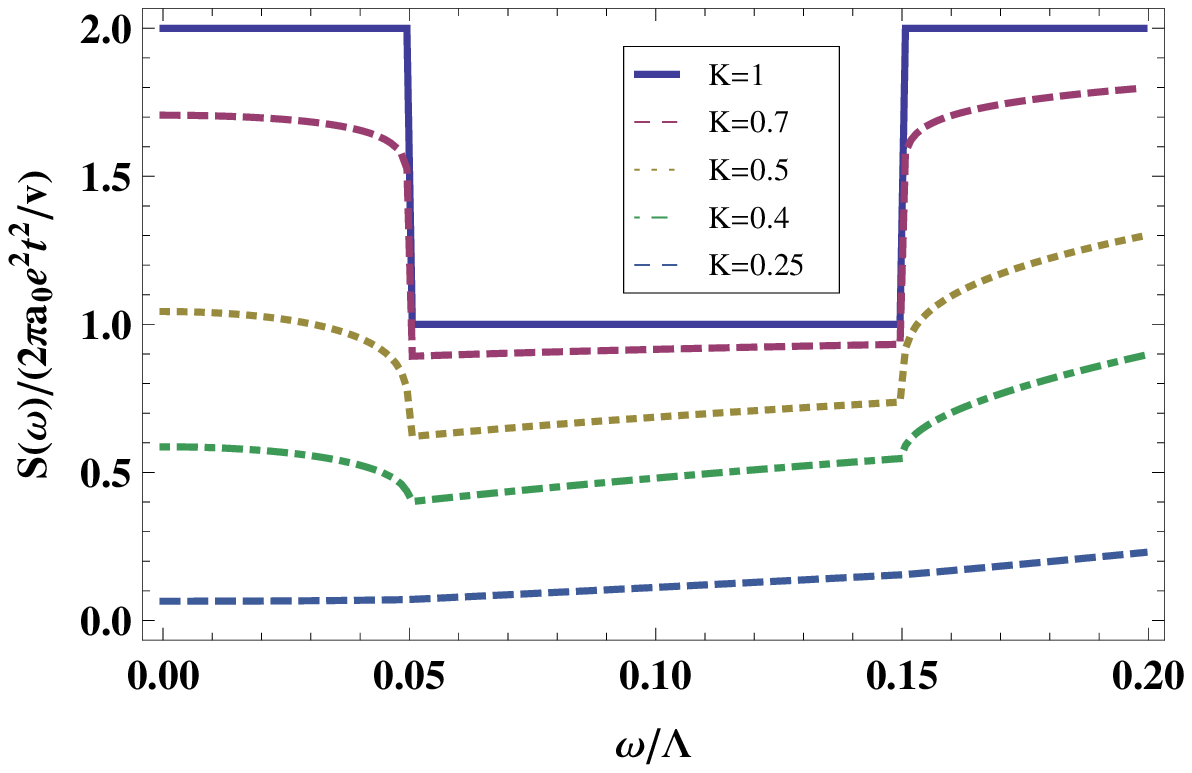}
\end{array}$
\end{center}
\caption{Noise for for different Luttinger parameters $K=1$ (blue line), $K=0.7$ (purple dashed line), $K=0.5$ (brown dotted line), $K=0.4$ (green dot-dashed
line), $K=0.25$ (light blue dashed line), and $\epsilon_d/\Lambda=-0.05$ in both figures. Left: Zero frequency noise $S(V,0)$ as a function of voltage. Right: $S(V,\omega)$ for fixed voltage with $eV/2\Lambda=0.1$ as a function of frequency.}
 \label{f3}
\end{figure*}

\begin{figure*}
\begin{center}$
\begin{array}{cc}
\includegraphics[width=1\columnwidth, clip]{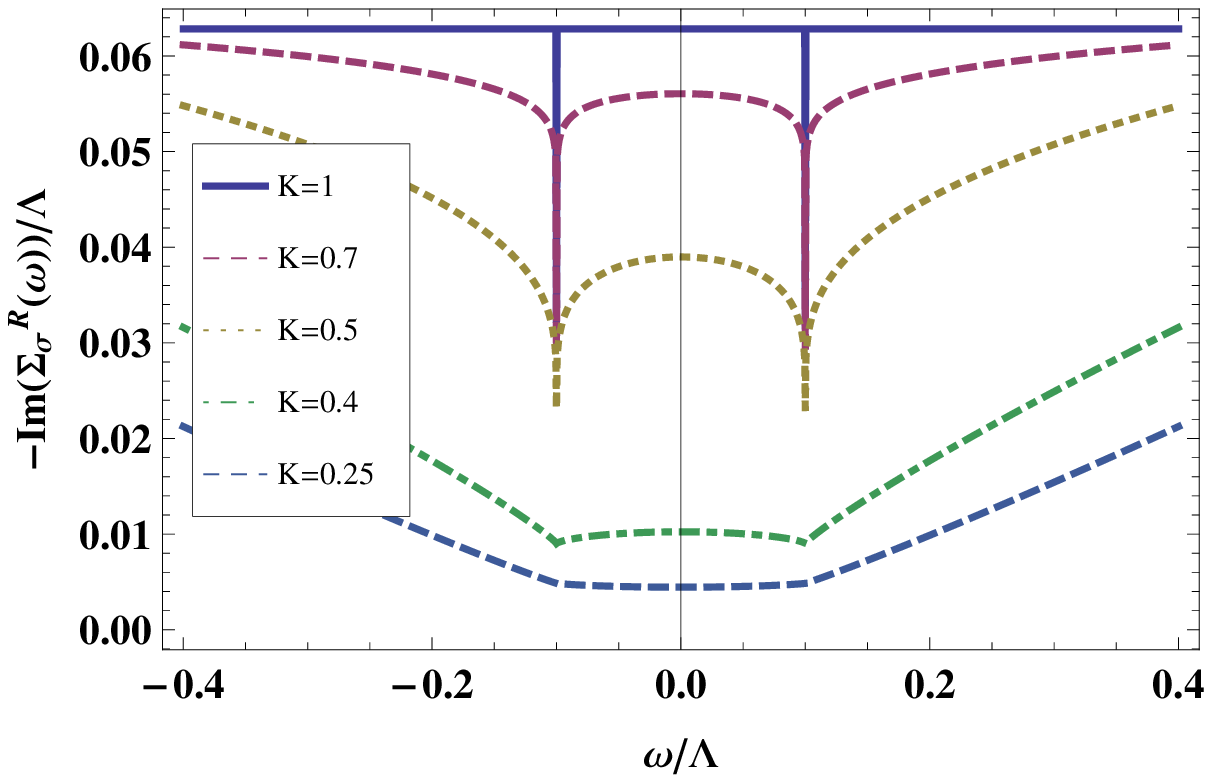}&
\includegraphics[width=1\columnwidth, clip]{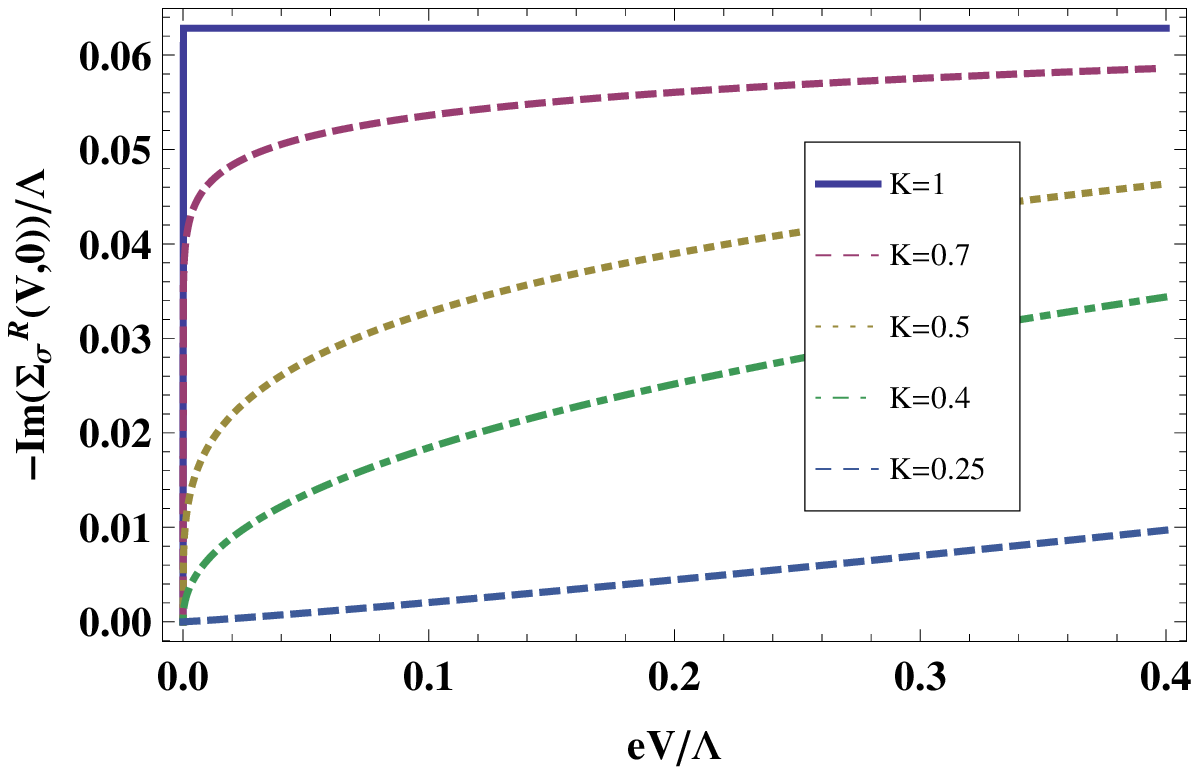}
\end{array}$
\end{center}
\caption{Lifetime for different Luttinger parameters $K=1$ (blue line), $K=0.7$ (purple dashed line), $K=0.5$ (brown dotted line), $K=0.4$ (green dot-dashed line), $K=0.25$ (light blue dashed line), and $t/\Lambda=0.1$ in both cases. Left: Lifetime for $|eV/\Lambda|=0.2$ as a function of frequency. Right: Lifetime for $\omega=0$ as a function of voltage.}
 \label{f4}
\end{figure*}

Note that though the hopping term $t$ here is relevant (at the tree-level)
in the metallic phase with a scaling
dimension $[t] = (K_c + K_s + 1/K_c + 1/K_s)/8 <1$ (see Eq.\ref{eq8}), our perturbative
calculation for U=0 is still controlled as the hopping term is treated
exactly due to the quadratic nature of our Hamiltonian in terms of the
impurity (dot) operator $d_{\sigma}$. Similar property can be found in other
systems, including the non-interacting single impurity Anderson model
\cite{hewson}, and the non-interacting pseudogap Anderson model\cite{Vojta}.  


For $\epsilon_d\neq 0$, i.e. dot level away from the equilibrium Fermi level of the two leads, as shown in the right figure of Fig.\ref{f1} the transition occurs at larger $K$ value for $0<K<1$ (say, showing insulating behavior at $K=0.5$ for $\epsilon_d/\Lambda=-0.1$ and $t/\Lambda=0.1$ shown as purple dot dashed line), indicating smaller repulsive interaction within the edge states would lead to insulating phase in this off resonance regime nearby equilibrium. When the dot level is far from the edge potentials, we can carry out an off resonance study by integrating out the dot electron state\cite{Chamon}, obtaining an effective single particle hopping term
\begin{eqnarray}
H_{int}\sim \frac{t_1 t_2^{\dagger}}{|\epsilon_d-\mu|}(\psi_{R\uparrow}(0)^{\dagger}\psi_{L\uparrow}(0)+\psi_{L\downarrow}(0)^{\dagger}\psi_{R\downarrow}(0))+h.c.
\end{eqnarray}
Here $\mu=(\mu_1+\mu_2)/2$ is the equilibrium Fermi level of the edge states and the projection is done assuming $eV=|\mu_1-\mu_2|\ll |\epsilon_d-\mu|$. This single particle tunneling current, as well as the current noise associated with it, has been analyzed by Lee et al. in Ref.~\onlinecite{Yuwen}. For $t=t_1=t_2$ the single particle tunneling current\cite{Yuwen} $I_t$ at zero temperature is
\begin{eqnarray}\label{offres}
I_t\propto \frac{e|t|^2}{|\epsilon_d-\mu|} V^{K+\frac{1}{K}-1}
\end{eqnarray}
which renders single particle tunneling conductance $dI_t/dV\propto V^{K+\frac{1}{K}-2}$ and the insulating phases occurs at $K\neq 1$. Thus for off-resonance case any nonzero interaction strength, attractive or repulsive, would lead to the insulating phase at zero bias. The power law behavior in Eq.(\ref{offres}) is consistent with our numerical results for off-resonance case shown in Fig.\ref{f1a0}. The inset of Fig.\ref{f1a0} shows a large portion of the conductance indeed
proportional to $V^{K+\frac{1}{K}-2}$. The upturn close to the resonance could be taken into account by including multi-particle tunneling events, as considered by
Kane and Fisher in Ref.~\onlinecite{Kane1} for two-particle tunneling process in the chiral Luttinger liquids or Teo and Kane in Ref.~\onlinecite{Teo} in the helical Luttinger liquids. The multi-particle tunneling events is a natural result from our analysis as we include all orders of perturbation through the dot self energy term. For example, the two-particle tunneling process\cite{Teo,Kane1,Yuwen} gives $dI_t/dV\propto V^{\left(\frac{4}{K}-2\right)}$ which is illustrated in the steepest dotted line in the inset of Fig. \ref{f1a0} for $K=0.2$ case. The critical Luttinger parameter for the two-particle tunneling
is obtained by the zero of the power in differential conductance which renders $K_{cr}=2$. The other critical value $K_{cr}=1/2$ corresponds to the conducting-insulating transition for spin current\cite{Teo, Hou}, which cannot be observed in this charge current analysis.

The fact that multi-particle tunneling processes show up nearby the resonance peak is understood as following. Due to the higher power in small parameter $t/|\epsilon_d-\mu|$ in perturbative expansion with the dot Green's function, the multi-particle tunneling process amplitude is diminishingly small for dot level $\epsilon_d$ away from the Fermi levels. Higher order terms become more important when $\epsilon_d$ is closer to one of the chemical potential in nonequilibrium regime.   


\subsection{Current noise and Lifetime of dot electron}
Similar to the chiral Luttinger liquids\cite{Chamon}, 
the metal-insulator transition driven by interaction within the helical Luttinger edge states can also be probed in the noise\cite{Schmidt,Simon, Yuwen} or phase sensitive measurement. The noise spectrum is given by the current current correlation $S(V,\omega)=\int dt e^{i\omega t}\langle \{ \Delta \hat{I}(t),\Delta \hat{I}(0)\}\rangle$. At zero temperature the lowest order perturbation of $S(V,\omega)$ is proportional to Fourier transform of $\langle\{ \hat{I}(t),\hat{I}(0)\}\rangle$ which is expressed as
\begin{eqnarray}\nonumber
&&\langle\{ \hat{I}(t),\hat{I}(0)\}\rangle \propto \sum_{\sigma}|t_1|^2\langle e^{-i\mu_1 t}\psi_{1\sigma}^{\dagger}(t)d_{\sigma}(t)d_{\sigma}^{\dagger}(0)\psi_{1\sigma}(0)\\\nonumber
&&+e^{i\mu_1 t}d_{\sigma}^{\dagger}(t)\psi_{1\sigma}(t)\psi_{1\sigma}^{\dagger}(0)d_{\sigma}(0)\rangle+|t_2|^2\langle e^{-i\mu_2 t}\psi_{2\sigma}^{\dagger}(t)d_{\sigma}(t)\\&&\times d_{\sigma}^{\dagger}(0)\psi_{2\sigma}(0)+e^{i\mu_2 t}d_{\sigma}^{\dagger}(t)\psi_{2\sigma}(t)\psi_{2\sigma}^{\dagger}(0)d_{\sigma}(0)\rangle
\end{eqnarray}
The zeroth order dot electron correlator in time domain is given by $\langle d^{\dagger}_{\sigma}(t)d_{\sigma}(0)\rangle=\theta(\epsilon_d)e^{-i\epsilon_d t}$ at zero temperature. Thus the lowest order $S(V,\omega)$ is 
\begin{eqnarray}\nonumber
&&S(V,\omega)\simeq \frac{2\pi e^2}{\Gamma(4\kappa)}\left(\frac{a_0}{v}\right)^{4\kappa}\Big\{|t_1|^2\big(|\omega-\mu_1+\epsilon_d|^{4\kappa-1}\\\nonumber&&\times\theta(\omega-\mu_1+\epsilon_d)+|\omega+\mu_1-\epsilon_d|^{4\kappa-1}\theta(-(\omega+\mu_1-\epsilon_d))\big)\\\nonumber
&&+|t_2|^2\big(|\omega-\mu_2+\epsilon_d|^{4\kappa-1}\theta(\omega-\mu_2+\epsilon_d)\\\label{12}&&+|\omega+\mu_2-\epsilon_d|^{4\kappa-1}
 \theta(-(\omega+\mu_2-\epsilon_d))\big)\Big\}
\end{eqnarray}
Here $\kappa=\frac{1}{8}\left(K+\frac{1}{K}\right)$. For lowest order perturbation $S(V,0)=ev\langle \hat{I}\rangle_0$ with $\langle \hat{I}\rangle_0$ denoting the lowest order current. The zero frequency noise as a function of voltage and fixed voltage noise as a function of frequency are plotted in the left and right figures of Fig. \ref{f3}. From left figure we see there is a discontinuity occurs at voltage $|eV/2|\sim |\epsilon_d|$ for $0.26<K<1$ and continuous curve for $0<K<0.26$ for repulsive interaction. This transition happens at the same critical value of Luttinger parameter for dot level in resonance with equilibrium Fermi level. The reason is that the continuity condition is determined by $\partial S(V,0)/\partial V$ which is proportional to differential conductance $dI/dV$ at the lowest order perturbation. Similar discontinuities are also found in the fixed voltage noise where $\partial S(V,\omega)/\partial \omega$ shows discontinuities at $\omega\sim |eV/2\pm \epsilon_d|$ for $0.26<K<1$.

The phase sensitive measurement such as Aharonov-Bohm ring measurement gives the information of dot lifetime defined as $-\Im\left(\Sigma_{\sigma}^{R}(V,\omega)\right)$. Following Appendix \ref{a1} with symmetric coupling $t_1=t_2$ we have
\begin{eqnarray}\label{lifetime}
-\Im\left(\Sigma_{\sigma}^{R}(V,\omega)\right)\propto \left(|\omega-\mu_1|^{4\kappa-1}+|\omega-\mu_2|^{4\kappa-1}\right)
\end{eqnarray}
As Eq.(\ref{lifetime}) shares similar power law behavior as Eq.(\ref{12}) for $\epsilon_d=0$ case, they show similar discontinuity behaviors.   
 The lifetime for fixed voltage or fixed frequency are plotted in Fig. \ref{f4}. For fixed voltage (left figure) we see again the dips structure at $\omega\sim \pm eV/2$ which is a symmetric function of frequency as we choose symmetrical couplings. For zero frequency we see discontinuity at $V\sim 0$ for $0.26<K<1$ in repulsive case. 

\begin{figure}
\includegraphics[width=1\columnwidth, clip]{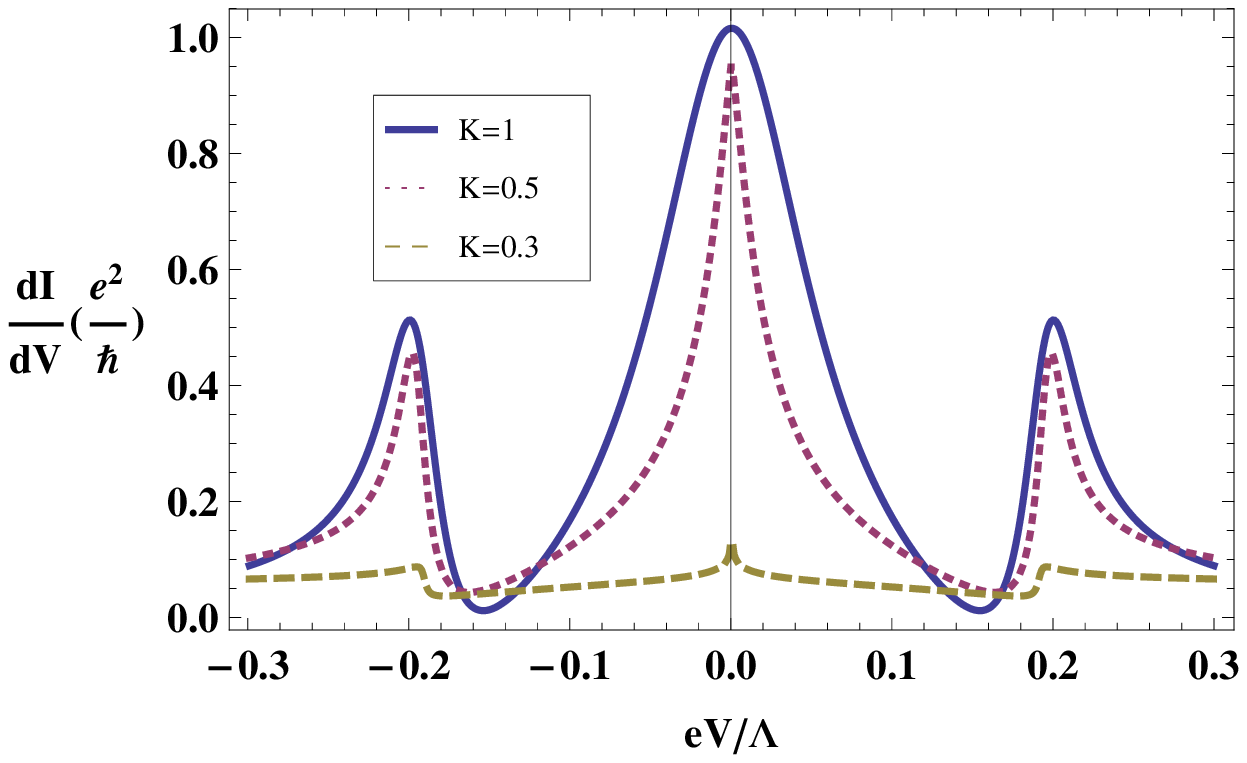}\\
\includegraphics[width=1\columnwidth, clip]{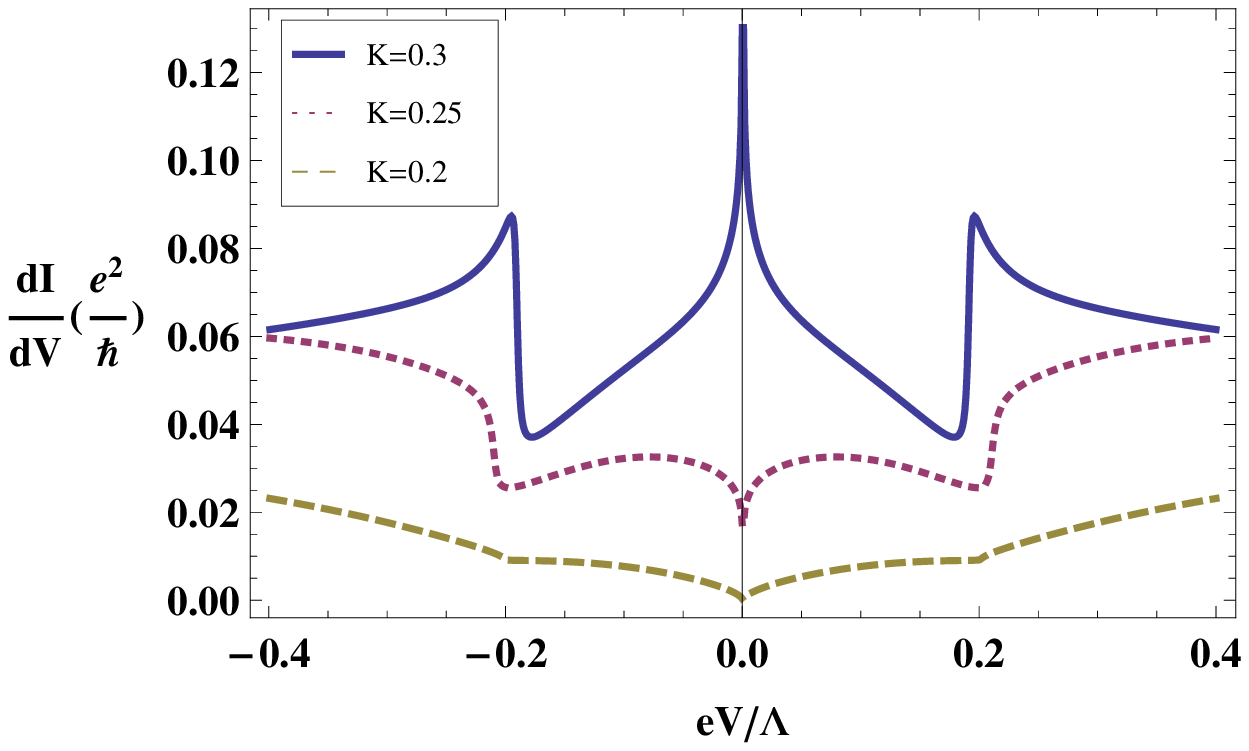}
\caption{Differential conductance v.s. source drain voltage for $t/\Lambda=0.075$, $\epsilon_d=0$, $U/\Lambda=0.1$, and $\mu_1=-\mu_2=eV/2$ with different interaction strengths within the edge states. Top: $K=1$ (blue solid line), $K=0.5$ (purple dotted line), $K=0.3$ (brown dashed line); Bottom: $K=0.3$ (blue solid line), $K=0.25$ (purple dotted line), $K=0.2$ (brown dashed line). }
 \label{f5}
\end{figure}

\section{Weakly interacting dot}\label{st4}
For small $U$ we use the equation of motion method to obtain the dot Green's function without coupling to the edge states. Then we take the interacting isolated dot Green's function as the unperturbed dot Green's function and add the self energy term in Eq.(\ref{se1}) as the approximated dressed dot Green's function. This
method is equivalent to make the Hartree-Fock approximation\cite{Haug} to the higher order correlation generated in the equation of motion approach with dot coupled to the edge states. Our approach, which is equivalent to that in Ref.~\onlinecite{hewson} for the single impurity Anderson model in the presence of weak on-site Coulomb interaction on the impurity ($U/t\ll 1$), is described as follows.

  The retarded Green's function of the isolated dot is
\begin{eqnarray*}
g^R_{\sigma\sigma}(t)=-i\theta(t)\langle\{d_\sigma(t),d_{\sigma}^{\dagger}(0)\}\rangle
\end{eqnarray*}
  By taking the time derivative and use the Heisenberg equation of motion $i \dot{d}_{\sigma}=\epsilon_d d_\sigma+U[d_\sigma,n_\sigma n_{\bar{\sigma}}]= \epsilon_d d_\sigma+U d_\sigma n_{\bar{\sigma}}$ to replace $\dot{d}_{\sigma}$ we obtain:
\begin{eqnarray}\label{g1}
i\frac{\partial g_{\sigma\sigma}^R(t)}{\partial t}=\delta(t)+\epsilon_d g_{\sigma\sigma}^R(t)+ U g_{\sigma\bar{\sigma}}(t)
\end{eqnarray} 
In Eq.(\ref{g1}) we define the two particle correlator $-i\theta(t)\langle\{d_\sigma(t) n_{\bar{\sigma}}(t),d_{\sigma}^{\dagger}(0)\}\rangle\equiv g_{\sigma\bar{\sigma}}(t)$. Take time derivative on $g_{\sigma\bar{\sigma}}(t)$ we get
\begin{eqnarray}\label{g2}
i\frac{\partial g_{\sigma\bar{\sigma}}(t)}{\partial t}=\delta(t)\langle n_{\bar{\sigma}}\rangle+\epsilon_d g_{\sigma\bar{\sigma}}(t)+U g_{\sigma\bar{\sigma}}(t)
\end{eqnarray}
Combining Eq.(\ref{g1}) and Eq.(\ref{g2}) and take the Fourier transform we obtain:
\begin{eqnarray}\label{g3}
g^R_{\sigma\sigma}(\omega)=\frac{\langle n_{\bar{\sigma}}\rangle}{\omega-\epsilon_d-U+i\delta}+\frac{1-\langle n_{\bar{\sigma}}\rangle}{\omega-\epsilon_d+i\delta}
\end{eqnarray}
The interacting dot Green's function connected to the edge states is approximated by
\begin{eqnarray}\label{g4}
G^R_{\sigma\sigma}(\omega)\simeq \frac{1}{(g^R_{\sigma\sigma}(\omega))^{-1}-\Sigma_{\sigma}^R(\omega)}
\end{eqnarray}
with the self energy given by Eq.(\ref{se1}). The $\langle n_{\bar{\sigma}}\rangle$ in Eq.(\ref{g3}) is obtained self consistently through numerical iterations by
\begin{eqnarray}
\langle n_{\bar{\sigma}}\rangle=\int\frac{d\omega}{2\pi i}G^{<}_{\bar{\sigma}\bar{\sigma}}(\omega)
\end{eqnarray}
Here $G^{<}_{\bar{\sigma}\bar{\sigma}}(\omega)=G^{R}_{\bar{\sigma}\bar{\sigma}}(\omega)\Sigma_{\bar{\sigma}}^{<}(\omega)G^{A}_{\bar{\sigma}\bar{\sigma}}(\omega)$.
The differential conductance is obtained by taking numerical derivative on the current and the result for different interaction strengths within the edge states 
is shown in Fig. \ref{f5}. 

\begin{figure}
\includegraphics[width=1\columnwidth, clip]{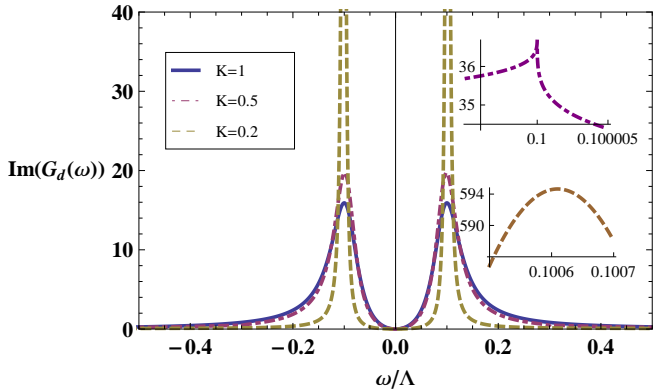}
\caption{Imaginary part of the nonequilibrium steady state dot Green's function for $t/\Lambda=0.1$, $\epsilon_d/\Lambda=-U/2\Lambda=-0.1$, and $\mu_1=-\mu_2=eV/2=0.1\Lambda$ with different interaction strengths within the edge states: $K=1$ (blue solid line), $K=0.5$ (purple dot dashed line), and $K=0.2$ (brown dashed line). Inset shows the blowup of $K=0.5$ (Upper inset) and $K=0.2$ (Lower inset) cases for regions $\omega\simeq |eV/2|$. Singular behavior for $0.26<K<1$ for repulsive interaction has the same root as singular behavior in the dot lifetime $-\Im(\Sigma_{\sigma}^R(\omega))$.}
 \label{f6}
\end{figure}

From Fig. \ref{f5} the metal-insulator transition nearby equilibrium occurs at the Luttinger parameter $K\simeq 0.26$ which is the same as the non-interacting dot case. The positions of the side peaks are at $\mu_1/\mu_2\sim \epsilon_d+U$ ($|eV/\Lambda|\sim 0.2$ in Fig. \ref{f5}) as a result of the charge fluctuations on the dot. This charge fluctuation side peaks become dips as the Luttinger parameter $K< 0.26$ and the overall magnitude as well as the width of the resonance peaks decreases with decreasing $K$ for $K<1$. The height of the side peaks are roughly half of the zero bias peak in Fig. \ref{f5} due to the alignment of Fermi seas as only one of them aligns with the dot level at finite voltage. 

In this equation of motion approach we ignore the contribution from 
higher order correlators and thus we cannot access the Kondo regime which occurs at larger $U$. This can be seen clearly from the lack of Kondo resonance peak in the dot density of state with zero source drain voltage, or the imaginary part of the dot Green's function in equilibrium. In Fig. \ref{f6} we plot the nonequilibrium steady state dot density of state, which can be probed by placing a scanning tunneling microscope(STM) tip on top of the dot. We find, as shown in the upper inset of Fig. \ref{f6} for $K=0.5$ case, singular behavior at $\omega\simeq \mu_1$ or $\mu_2$ for Luttinger parameter $K$ greater than $0.26$ for repulsive cases while no singular behavior is seen for $K<0.26$. This is similar to the situation for lifetime discussed in Fig. \ref{f4}. The width of the charge resonance peaks centered around $\omega\sim\epsilon_d$ and $\epsilon_d+U$ decreases with increasing interaction strengths while increases with increasing source drain voltage. In the off resonance regime where $|\omega-\epsilon_d|/(t^2\frac{a}{\hbar v})\gg 1$ and $|\omega-\epsilon_d-U|/(t^2\frac{a}{\hbar v})\gg 1$
the dot density of state goes as $|\omega|^{4\kappa-3}$ as expected from $Im(G_d^R(\omega))\sim \sum_{\sigma}Im(\Sigma_{\sigma}^R(\omega))/|\omega|^2 \propto| \omega|^{4\kappa-3}$ in the off resonance regime. 

For the interacting quantum dot system we may tune the Luttinger parameters of the edge states to realize the transition from one-channel to two-channel Kondo physics\cite{Law,Chung}. The required Luttinger parameter for repulsive case fall between $1$ and $1/2$ with higher order renormalization group analysis\cite{Chung}, indicating a weaker repulsive interaction is needed to see this transition in the Kondo regime (corresponding to $\epsilon_d+U/2=0$ and $\epsilon_d\rightarrow -\infty$ to suppress charge fluctuations with particle-hole symmetry in the Anderson impurity model) compared with dot level in resonance with the equilibrium Fermi surface at zero bias. This is again consistent with the results in non-interacting dot as physics of the one channel to two channel Kondo transition is closely related to the metal to insulating transition for this two leads setup.

\section{Conclusions}\label{st5}
In this paper, we consider the non-equilibrium charge current through a non-interacting or weakly interacting quantum dot connected by two helical Luttinger liquids leads. For non-interacting leads the current is obtained exactly by inclusion of all orders of Keldysh perturbations in the couplings between leads and dot and the interacting leads Green's functions solved by standard bosonization. We find stronger interactions, with Luttinger parameter $K<0.26$ in the repulsive or $K>3.73$ in the attractive case, is required to see the metallic to insulating transition at zero bias when the dot level is in resonance with the equilibrium Fermi level. For off-resonance dot level the transition occurs at weaker interaction strengths, reaching noninteracting limit if dot level were far away from the zero bias Fermi surface. This is consistent with the naive expectation as the metal to insulator transition is more susceptible to the interaction
of the leads if the state itself were already close to the insulating phase. Within the lowest perturbation theory with dot in resonance with the equilibrium Fermi level, other physical quantities such as current noise, nonequilibrium lifetime, and nonequilibrium density of state of the dot electron also show sharp to smooth transitions at $K\sim 0.26$, consistent with the scaling dimension analysis. 
    
For weakly interacting dot the dot Green's function is obtained perturbatively with Hatree-Fock approximations. The differential conductance is again obtained by taking numerical derivative on the current-voltage relation. We find similar QPT as the non-interacting dot case and the charge fluctuation side peak shows similar behaviors as the zero bias peak for dot level in resonance with the zero bias Fermi level. For dot level tuned to the Kondo regime the required interaction strengths in the edge states to see the transition from one-channel to two-channel Kondo is weaker\cite{Law,Chung}, making the system ideal for observing the two-channel Kondo physics.

Finally let us comment on the experimental feasibility of realizing this quantum phase transition and current experimental results realizing the edge state transport of QSHI on the HgTe/CdTe quantum well devices. The rough estimate on the Luttinger parameter $K$ is given by $K^2\sim (1+\frac{U}{2\epsilon_F})^{-1}$, where $\epsilon_F$ is the Fermi energy and $U$ is the characteristic Coulomb energy of the edge states\cite{Kane2}. For a HgTe/CdTe quantum well the Coulomb energy can be controlled by the width $w$ of the well\cite{Hou} (with $U\sim e^2/w$) or placing a substrate which changes the dielectric parameters on the edge state (with $U\sim e^2/\epsilon a_0$, $\epsilon$ being dielectric constant). The Luttinger parameter is estimated to be $K\simeq 0.55$ for HgTe/CdTe experimental devices\cite{Konig,Molenkamp} and reaches $K\simeq 0.35$ for a similar device without the top gate\cite{David}. This value is close to the large repulsive  interaction strength we need ($K\sim 0.26$) as well as the regime of $K<1/4$ required for the "Luttinger liquid insulator" or instability threshold of the edge state\cite{Wu,Xu}. Thus the edge states leads are well established and fabricating a quantum dot through depositing or e-beam lithography on the QSHI junction devices in principal can realize the experimental setup mentioned in this article.  

 Furthermore, in Ref.~\onlinecite{David} M. K{\"o}nig et al. performed spatially resolved study of backscattering in the Quantum Spin Hall state using scanning gate microscopy. They find backscattering rate at well localized sites can be tuned by the backgate. By treating the quantum dot as an impurity site at one edge of the device we can measure the relation between the backscattering rate and the Luttinger parameter $K$. Our results suggest that  
the coupling between the impurity and edge states decreases with increasing interaction strengths, which can be probed by the backscattering rate measurement.
In principal the interaction strengths of two edges need not be the same and the $K_{cr}$ would change given this asymmetry in the real experimental setup. The
qualitative features we discuss in this article should still be valid nearby the metal to insulator transition. 
  
\section*{Acknowledgment}
 S.-P. acknowledges useful comments from Yu-Wen Lee, Yu-Li Lee, and Thomas Schmidt, and the financial support from the National Center for Theoretical Sciences in Taiwan. CHC acknowledges the support from the NSC grant No.101-2628-M-009-001-MY3,
the MOE-ATU program, the CTS of NCTU, the NCTS of Taiwan, R.O.C.

\appendix
\section{Evaluation of correlation functions}\label{a1}
The bare correlation function $-i\langle T_c\{\psi_{R/L\sigma}(\tau_1)\psi^{\dagger}_{R/L\sigma}(\tau_2)\}\rangle$ involves evaluating 
 $-i\langle T_c\{e^{i\sqrt{4\pi}(\phi_{R/L\sigma}(\tau_1)-\phi_{R/L\sigma}(\tau'))}\eta_{R/L\sigma}(\tau_1)\eta_{R/L\sigma}(\tau_2) \}\rangle$. The bare action $S_0$ associated with helical leads Hamiltonian $H_0$ is
\begin{eqnarray*} 
&&-S_0=\int_0^\beta d\tau \int dx \Big\{\sum_\alpha [i\nabla\Theta_{\alpha}(x,\tau)\partial_{\tau}\Phi_{\alpha}(x,\tau)\\
&&-\frac{v_\alpha}{2}(K_\alpha(\nabla\Theta_{\alpha})^2+\frac{1}{K_\alpha}(\nabla\Phi_{\alpha})^2 )]+\sum_\sigma d^{\dagger}_\sigma(\partial_\tau-\epsilon_d)d_\sigma\Big\}
\end{eqnarray*}
 Here $r_j^{\alpha}=(x_j,v_\alpha \tau_j)$ and we denote ${\bf q_\alpha}=(k,\omega_n/v_\alpha)$ as its Fourier momentum for later use.
Expressing $\phi_{R\sigma}=(\Phi_c+sgn(\sigma)\Phi_s-\Theta_c-sgn(\sigma)\Theta_s)/2\sqrt{2}$ and $\phi_{L\sigma}=(\Phi_c+sgn(\sigma)\Phi_s+\Theta_c+sgn(\sigma)\Theta_s)/2\sqrt{2}$ we get the general form of correlation
function as
\begin{eqnarray}\nonumber
I&=&\langle T_c\{\prod_{\alpha,j} e^{i( A_j^{\alpha}\Phi_{\alpha}(r_j^{\alpha})+B_j^{\alpha}\Theta_{\alpha}(r_j^{\alpha}))}\}\rangle\\\label{ea1}
&=&e^{-\frac{1}{2}\langle T_c[\sum_{\alpha,j}( A_j^{\alpha}\Phi_{\alpha}(r_j^{\alpha})+B_j^{\alpha}\Theta_{\alpha}(r_j^{\alpha}))]^2\rangle}
\end{eqnarray}
The effect of Klein factor $-i\langle T_c \eta_{R/L\sigma}(\tau_1)\eta_{R/L\sigma}(\tau_2)\rangle$, valued at $\pm i$ depending on the ordering, will be included in the $\langle\Phi_{\alpha}(r_j^{\alpha})\Theta_{\alpha}(r_j^{\alpha})\rangle$ term later.
 The second line of Eq.(\ref{ea1}) has used the quadratic nature of the bosonic field $\Phi_{c,s}$ and $\Theta_{c,s}$ in the action $S_0$.
 
 From the Fourier component and choose $\tau_2$ on the top and $\tau_1$ on the bottom of the Keldysh contour: 
\begin{eqnarray*}
&&\langle \Phi_{\alpha}(r_1)\Phi_{\alpha'}(r_2)\rangle=\frac{\delta_{\alpha,\alpha'}}{\beta\Omega}\sum_{{\bf q_\alpha}}\langle \Phi_{\alpha}({\bf q_\alpha})\Phi_{\alpha}(-{\bf q_\alpha})\rangle e^{i {\bf q_\alpha}(r_1-r_2)}\\
&&=\frac{\delta_{\alpha,\alpha'}}{\beta}\sum_{\omega_n}\int \frac{dk}{2\pi}\frac{ K_\alpha}{\omega_n^2/v_\alpha+v_\alpha k^2}e^{i(kx-\omega_n \tau)}\\
&&=\delta_{\alpha,\alpha'}K_\alpha\Big\{\int \frac{dk}{8\pi k}e^{-a_0|k|}[-f_B(-v_\alpha k)\theta(-k)e^{ikx+v_\alpha k\tau}\\
&&+f_B(v_\alpha k)\theta(k)e^{ikx-v_\alpha k\tau}]+\int \frac{dk}{8\pi k}e^{-a_0|k|}[\theta(k)e^{ikx-v_\alpha k\tau}\\
&&-\theta(-k)e^{ikx+v_\alpha k\tau}]\Big\}\\
&&=\delta_{\alpha,\alpha'}K_\alpha \Big\{\int_0^{\infty} \frac{dk}{4\pi k}e^{-a_0|k|}f_B(v_\alpha k)\cos(kx)e^{-v_\alpha k\tau}\\
&&+\int_0^{\infty} \frac{dk}{4\pi k}e^{-a_0|k|}\cos(kx)e^{-v_\alpha k\tau}\Big\}\\
&&\langle \Theta_{\alpha}(r_1)\Theta_{\alpha'}(r_2)\rangle=\frac{\delta_{\alpha,\alpha'}}{\beta}\sum_{\omega_n}\int \frac{dk}{2\pi}\frac{1/ K_\alpha}{\omega_n^2/v_\alpha+v_\alpha k^2}e^{i(kx-\omega_n \tau)}\\
&&=\langle \Phi_{\alpha}(r_1)\Phi_{\alpha'}(r_2)\rangle/K_\alpha^2\\
&&\langle \Phi_{\alpha}(r_1)\Theta_{\alpha'}(r_2)\rangle=\frac{\delta_{\alpha,\alpha'}}{\beta}\sum_{\omega_n}\int \frac{dk}{2\pi}\frac{-i\omega_n}{k(\omega_n^2+v_\alpha^2 k^2)}e^{i(kx-\omega_n \tau)}\\
&&=\delta_{\alpha,\alpha'}\Big\{\int \frac{dk}{8\pi k}e^{-a_0|k|}[f_B(-v_\alpha k)\theta(-k)e^{ikx+v_\alpha k\tau}\\
&&+f_B(v_\alpha k)\theta(k)e^{ikx-v_\alpha k\tau}]+\int \frac{dk}{8\pi k}e^{-a_0|k|}[\theta(k)e^{ikx-v_\alpha k\tau}\\
&&+\theta(-k)e^{ikx+v_\alpha k\tau}]\Big\}\\
&&=\delta_{\alpha,\alpha'} \Big\{\int_0^{\infty} \frac{dk}{4\pi k}e^{-a_0|k|}f_B(v_\alpha k)i\sin(kx)e^{-v_\alpha k\tau}\\
&&+\int_0^{\infty} \frac{dk}{4\pi k}e^{-a_0|k|}i\sin(kx)e^{-v_\alpha k\tau}\Big\}
\end{eqnarray*}
Here $f_B(x)=1/(e^{\beta x}-1)$ is the Bose-Einstein distribution function and $\beta=1/k_B T$ is the inverse temperature.
 At zero temperature $f_B(x)=-\theta(-x)$. We carry out the momentum integral with $1/a_0$ as the high energy cutoff (or equivalently putting in $e^{-a_0 k}$ in the momentum integral) and analytically continue the imaginary time to real time component to obtain
\begin{eqnarray}\nonumber
&&\frac{1}{K_\alpha}\langle \Phi_{\alpha}(r_1)\Phi_{\alpha'}(r_2)\rangle=\frac{-1}{4\pi}\ln\left[\frac{x^2+(a_0+iv_\alpha t)^2}{a_0^2}\right]\\
&&  \equiv F_\alpha^{(1)-+}(t,x)\\\nonumber
&&\langle \Phi_{\alpha}(r_1)\Theta_{\alpha'}(r_2)\rangle=\frac{-1}{4\pi}\ln\left[\frac{a_0+iv_\alpha t-i x}{a_0+iv_\alpha t+i x}\right]\\
&&\equiv F_\alpha^{(2)-+}(t,x)
\end{eqnarray}
Here $t=t_1-t_2$ and $x=x_1-x_2$.
For $t_2$ on the bottom and $t_1$ on the top branch of Keldysh contour we substitute $x\rightarrow -x$ and $t_1\leftrightarrow t_2$ to get
\begin{eqnarray}
&&F_\alpha^{(1)+-}(t,x)=\frac{-1}{4\pi}\ln\left[\frac{x^2+(a_0-iv_\alpha t)^2}{a_0^2}\right]\\
&&F_\alpha^{(2)+-}(t,x)=\frac{-1}{4\pi}\ln\left[\frac{a_0-iv_\alpha t+i x}{a_0-iv_\alpha t-i x}\right]
\end{eqnarray}
For both $t_2$ and $t_1$ on the top branch, or time ordered branch, we get
\begin{eqnarray}\nonumber
F_\alpha^{(1)++}(t,x)&=&\theta(t)F_\alpha^{(1)-+}(t,x)+\theta(-t)F_\alpha^{(1)+-}(t,x)\\\label{a6}
&=&\frac{-1}{4\pi}\ln\left[\frac{x^2+(a_0+iv_\alpha |t|)^2}{a_0^2}\right]\\\nonumber
F_\alpha^{(2)++}(t,x)&=&\theta(t)F_\alpha^{(2)-+}(t,x)+\theta(-t)F_\alpha^{(2)+-}(t,x)\\\label{a7}
&=&\frac{-1}{4\pi}\ln\left[\frac{a_0+iv_\alpha |t|-i sgn[t] x}{a_0+iv_\alpha |t|+i sgn[t]x}\right]
\end{eqnarray}
Similarly for anti-time ordered $F_\alpha^{(1)--}(t,x)$ and $F_\alpha^{(2)--}(t,x)$, obtained by $\theta(t)\leftrightarrow\theta(-t)$ in Eq.(\ref{a6}) and Eq.(\ref{a7}), are
\begin{eqnarray}
F_\alpha^{(1)--}(t,x)&=&\frac{-1}{4\pi}\ln\left[\frac{x^2+(a_0-iv_\alpha |t|)^2}{a_0^2}\right]\\
F_\alpha^{(2)--}(t,x)&=&\frac{-1}{4\pi}\ln\left[\frac{a_0-iv_\alpha |t|+i sgn[t] x}{a_0-iv_\alpha |t|-i sgn[t]x}\right]
\end{eqnarray}
We absorb the effect of Klein factor $-i\langle T_c \eta_{R/L\sigma}(\tau_1)\eta_{R/L\sigma}(\tau_2)\rangle$ by introducing $\tilde{F}_\alpha^{(2)++/--}(t,x)=F_\alpha^{(2)}(t,x)\pm sgn[t]\frac{i}{4}$ and $\tilde{F}_\alpha^{(2)+-/-+}(t,x)=F_\alpha^{(2)}(t,x)\pm \frac{i}{4}$.
The general form of $G_{\psi_{j,\sigma}}(\omega)$ at zero temperature is
\begin{eqnarray}\nonumber
&&G_{\psi_{j,\sigma}}(\omega)=\int_{-\infty}^{\infty}dt e^{i(\omega-\mu_{j,\sigma})t}e^{\frac{\pi}{2}\left(K_c+\frac{1}{K_c}\right)F_c^{(1)}(t,0)}\\\label{ggr}
&&\times e^{\frac{\pi}{2}\left(K_s+\frac{1}{K_s}\right)F_s^{(1)}(t,0)}e^{\pi\left(\tilde{F}_c^{(2)}(t,0)+\tilde{F}_s^{(2)}(t,0)\right)}
\end{eqnarray}
To compute Eq.(\ref{ggr}) let us first define $P_\alpha(\omega)$ as
\begin{eqnarray}\nonumber
&&\frac{1}{v_\alpha}P_{\alpha}(\frac{\omega}{v_\alpha})\\\nonumber&&\equiv\frac{1}{v_\alpha}\int_{-\infty}^{\infty}dt e^{i\omega \frac{t}{v_\alpha}}e^{\pi\left(4\kappa_\alpha F_\alpha^{(1)}(\frac{t}{v_\alpha},0)+\tilde{F}_\alpha^{(2)}(\frac{t}{v_\alpha},0)\right)}\\
&&\kappa_\alpha\equiv\frac{1}{8}\left(K_\alpha+\frac{1}{K_\alpha}\right)
\end{eqnarray}
From above we get the four Keldysh contour orders $P_{\alpha}(\frac{\omega}{v_\alpha})$ as     
\begin{eqnarray*}
&&P^{++}_{\alpha}(\frac{\omega}{v_\alpha})=a_0^{2\kappa_\alpha}\int_{-\infty}^{\infty}dt\frac{e^{i\omega \frac{t}{v_\alpha}}e^{i\frac{\pi}{4}sgn[t]}}{(a_0+i|t|)^{2\kappa_\alpha}}      \\
&&P^{--}_{\alpha}(\frac{\omega}{v_\alpha})=a_0^{2\kappa_\alpha}\int_{-\infty}^{\infty}dt\frac{e^{i\omega \frac{t}{v_\alpha}}e^{-i\frac{\pi}{4}sgn[t]}}{(a_0-i|t|)^{2\kappa_\alpha}}       \\ 
&&P^{+-}_{\alpha}(\frac{\omega}{v_\alpha})=a_0^{2\kappa_\alpha}\int_{-\infty}^{\infty}dt\frac{e^{i\omega \frac{t}{v_\alpha}}e^{i\frac{\pi}{4}}}{(a_0-it)^{2\kappa_\alpha}}       \\ 
&&P^{-+}_{\alpha}(\frac{\omega}{v_\alpha})=a_0^{2\kappa_\alpha}\int_{-\infty}^{\infty}dt\frac{e^{i\omega \frac{t}{v_\alpha}}e^{-i\frac{\pi}{4}}}{(a_0+it)^{2\kappa_\alpha}} 
\end{eqnarray*}
Note that $P^{--}_{\alpha}(\frac{\omega}{v_\alpha})=[P^{++}_{\alpha}(-\frac{\omega}{v_\alpha})]^{\ast}$ and $P^{+-}_{\alpha}(\frac{\omega}{v_\alpha})=[P^{-+}_{\alpha}(-\frac{\omega}{v_\alpha})]^{\ast}$ and we only need to evaluate $P_\alpha^{++}$ and $P_\alpha^{+-}$.
For $P^{++}_{\alpha}(\omega)$ with $\omega>0$ we choose the following contour integral 
\begin{figure}
\includegraphics[width=.5\columnwidth, clip]{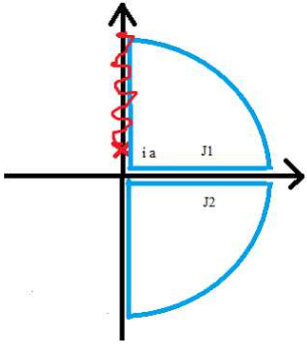}
\caption{Contour chosen for evaluating $P^{++}_{\alpha}(\omega)$ in the complex plane. Red curvy line is the branch cut. In Eq.(\ref{p1}) $J_{1\alpha}(\omega)$ is evaluated in the real axis of the upper corner and $J_{2\alpha}(\omega)$ is evaluated in the real axis of the lower corner.}
 \label{fc}
\end{figure}

\begin{eqnarray}\nonumber
&&P^{++}_{\alpha}(\omega)\\\nonumber&&=a_0^{2\kappa_\alpha}\left(\int_0^{\infty}dt\frac{e^{i(\omega t-\frac{\pi}{4})}}{(a_0+i t)^{2\kappa_\alpha}}+\int_0^{\infty}dt\frac{e^{i(-\omega t+\frac{\pi}{4})}}{(a_0+i t)^{2\kappa_\alpha}}\right)\\\label{p1}
&&=a_0^{2\kappa_\alpha}\left( J_{1\alpha}(\omega)+J_{2\alpha}(\omega)\right)\\\nonumber
&&J_{1\alpha}(\omega)=ie^{-2\pi i \kappa_\alpha}\Gamma(1-2\kappa_\alpha)\omega^{2\kappa_\alpha-1}e^{-a_0\omega-i\frac{\pi}{4}}\\\nonumber
&&J_{2\alpha}(\omega)=-ie^{a_0\omega}\omega^{2\kappa_\alpha-1}\Gamma(1-2\kappa_\alpha,a_0\omega)e^{i\frac{\pi}{4}}
\end{eqnarray}
Here $\Gamma(K,z)\equiv \int_{z}^{\infty}t^{K-1}e^{-t}dt$ is the incomplete Gamma function and $\Gamma(K,0)=\Gamma(K)$. For $a_0\omega\rightarrow 0$ and $\omega>0$ we simplify Eq.(\ref{p1}) as
\begin{eqnarray*}
P^{++}_{\alpha}(\omega)\simeq a_0^{2\kappa_\alpha}(ie^{-i(2\pi \kappa_\alpha+\frac{\pi}{4})}-ie^{i\frac{\pi}{4}})\Gamma(1-2\kappa_\alpha)\omega^{2\kappa_\alpha-1}
\end{eqnarray*}
For $\omega<0$ we choose left semi circle to perform the integral and we get
\begin{eqnarray*}
P^{++}_{\alpha}(\omega)\simeq a_0^{2\kappa_\alpha}(ie^{-i(2\pi \kappa_\alpha-\frac{\pi}{4})}-ie^{-i\frac{\pi}{4}})\Gamma(1-2\kappa_\alpha)(-\omega)^{2\kappa_\alpha-1}
\end{eqnarray*}
Since $P^{++}_{\alpha}(\omega)$ is an even function of $\omega$ and $P^{--}_{\alpha}(\omega)=[P^{++}_{\alpha}(-\omega)]^{\ast}$ we have
\begin{eqnarray*}
P^{++}_{\alpha}(\omega)&\simeq& a_0^{2\kappa_\alpha}i(e^{-i(2\pi \kappa_\alpha+\frac{sgn(\omega)\pi}{4})}-e^{i\frac{sgn(\omega)\pi}{4}})\\&\times&\Gamma(1-2\kappa_\alpha)|\omega|^{2\kappa_\alpha-1}\\
P^{--}_{\alpha}(\omega)&\simeq&- a_0^{2\kappa_\alpha}i(e^{i(2\pi \kappa_\alpha-\frac{sgn(\omega)\pi}{4})}-e^{i\frac{sgn(\omega)\pi}{4}})\\&\times&\Gamma(1-2\kappa_\alpha)|\omega|^{2\kappa_\alpha-1}
\end{eqnarray*}
For $P^{+-}_{\alpha}(\omega)$ we choose a different contour to perform the complex integrals and we get
\begin{eqnarray*}
&&P^{+-}_{\alpha}(\omega)=(-ia_0^{2\kappa_\alpha}e^{i2\pi (\kappa_\alpha+\frac{1}{8})}e^{-|\omega|a_0}\int_{\infty}^0 dr r^{-2\kappa_\alpha}e^{-|\omega|r}\\
&&+ia_0^{2\kappa_\alpha}e^{-i2\pi (\kappa_\alpha-\frac{1}{8})}e^{-|\omega|a}\int_{\infty}^0 dr r^{-2\kappa_\alpha}e^{-|\omega|r})\theta(-\omega)\\
&&=2a_0^{2\kappa_\alpha}\sin(2\pi \kappa_\alpha)e^{i\frac{\pi}{4}}|\omega|^{2\kappa_{\alpha}-1}e^{-|\omega| a_0}\Gamma(1-2\kappa_\alpha)\theta(-\omega)
\end{eqnarray*}
Thus for $a_0\omega\rightarrow 0$ we have
\begin{eqnarray*}
&&P^{+-}_{\alpha}(\omega)\simeq 2\sin(2\pi \kappa_\alpha)e^{i\frac{\pi}{4}}\frac{|a_0\omega|^{2\kappa_{\alpha}}}{|\omega|}\Gamma(1-2\kappa_\alpha)\theta(-\omega)\\
&&P^{-+}_{\alpha}(\omega)\simeq 2\sin(2\pi \kappa_\alpha)e^{-i\frac{\pi}{4}}\frac{|a_0\omega|^{2\kappa_{\alpha}}}{|\omega|}\Gamma(1-2\kappa_\alpha)\theta(\omega)
\end{eqnarray*}

Using expressions above we rewrite Eq.(\ref{ggr}) as
\begin{eqnarray*}
G_{\psi_{j,\sigma}}^{\mu\mu'}(\omega)=\int_{-\infty}^{\infty}\frac{d\nu}{2\pi v_s v_c}P_c^{\mu\mu'}\left(\frac{\nu}{v_c}\right)P_s^{\mu\mu'}\left(\frac{\omega-\mu_{j,\sigma}-\nu}{v_s}\right)
\end{eqnarray*}
with $\mu$, $\mu'$ denoting $\pm$. Define $P_{\mu,\mu'}(\omega)$ as
\begin{eqnarray}\label{a9}
P_{\mu,\mu'}(\omega)\equiv \frac{1}{v_s v_c}\int \frac{d\nu}{2\pi}P_c^{\mu,\mu'}(\frac{\nu}{v_c})P_s^{\mu,\mu'}(\frac{\omega-\nu}{v_s})
\end{eqnarray}
we see that $G_{\psi_{j,\sigma}}^{\mu\mu'}(\omega)=P_{\mu,\mu'}(\omega-\mu_{j,\sigma})$. Evaluation of Eq.(\ref{a9}) involves the following three integrals:
\begin{eqnarray*}
&&\int_{-\infty}^{\infty}d\nu |\nu|^{2\kappa_c-1}|\omega-\nu|^{2\kappa_s-1}e^{-\frac{|\nu|a_{0}}{v_c}-\frac{|\omega-\nu|a_{0}}{v_s}}\\
&&\simeq \frac{\Gamma(2\kappa_c)\Gamma(2\kappa_s)}{\Gamma(2\kappa_c+2\kappa_s)}h(\kappa_c,\kappa_s)\Big(|\omega|^{2\kappa_c+2\kappa_s-1}\theta(\omega)\\&&+|\omega|^{2\kappa_c+2\kappa_s-1}\theta(-\omega)\Big)+O(a_0^{1-2\kappa_s-2\kappa_c})\\
&&\int_{-\infty}^{\infty}d\nu |\nu|^{2\kappa_c-1}|\omega-\nu|^{2\kappa_s-1}\theta(-\nu)\theta(\nu-\omega)e^{-\frac{|\nu|a_0}{v_c}-\frac{|\omega-\nu|a_0}{v_s}}\\
&&\simeq\frac{\Gamma(2\kappa_c)\Gamma(2\kappa_s)}{\Gamma(2\kappa_c+2\kappa_s)}|\omega|^{2\kappa_c+2\kappa_s-1}\theta(-\omega)\\
&&\int_{-\infty}^{\infty}d\nu |\nu|^{2\kappa_c-1}|\omega-\nu|^{2\kappa_s-1}\theta(\nu)\theta(\omega-\nu)e^{-\frac{|\nu|a_0}{v_c}-\frac{|\omega-\nu|a_0}{v_s}}\\
&&\simeq\frac{\Gamma(2\kappa_c)\Gamma(2\kappa_s)}{\Gamma(2\kappa_c+2\kappa_s)}|\omega|^{2\kappa_c+2\kappa_s-1}\theta(\omega)
\end{eqnarray*}
Here $h(\kappa_c,\kappa_s)=1+\frac{\Gamma(1-2\kappa_c-2\kappa_s)\Gamma(2\kappa_c+2\kappa_s)}{\Gamma(2\kappa_c)\Gamma(1-2\kappa_c)}+\frac{\Gamma(1-2\kappa_c-2\kappa_s)\Gamma(2\kappa_c+2\kappa_s)}{\Gamma(2\kappa_s)\Gamma(1-2\kappa_s)}$. Again we have taken $a_0\rightarrow 0$ in the computation and the divergent term proportional to $a_0^{1-2\kappa_s-2\kappa_c}$ for $1-2\kappa_s-2\kappa_c<0$ in the first expression above cancel each other in the evaluation of current. Since $v_s=v_c=v$ and $K_c=1/K_s=K$ we have $\kappa_c=\kappa_s\equiv\kappa=\frac{1}{8}(K+1/K)$ and collecting the above results we get
\begin{eqnarray}
G_{\psi_{j,\sigma}}^{++}(\omega)&=&\frac{ a_0^{4\kappa}}{2\pi v^{4\kappa}}\frac{\Gamma(2\kappa)^2}{\Gamma(4\kappa)}|\omega-\mu_{j,\sigma}|^{4\kappa-1}\\\nonumber
&\times&\left(\tilde{h}(\kappa)\theta(\omega-\mu_{j,\sigma})-\tilde{h}(\kappa)\theta(\mu_{j,\sigma}-\omega)\right)\\\nonumber
G_{\psi_{j,\sigma}}^{--}(\omega)&=&\frac{ a_0^{4\kappa}}{2\pi v^{4\kappa}}\frac{\Gamma(2\kappa)^2}{\Gamma(4\kappa)}|\omega-\mu_{j,\sigma}|^{4\kappa-1}\\\nonumber
&\times&\left(-\tilde{h}^{\ast}(\kappa)\theta(\omega-\mu_{j,\sigma})+\tilde{h}^{\ast}(\kappa)\theta(\mu_{j,\sigma}-\omega)\right)\\\nonumber
G_{\psi_{j,\sigma}}^{+-}(\omega)&=&\frac{2\pi a_0^{4\kappa}}{v^{4\kappa}}\frac{i}{\Gamma(4\kappa)}|\omega-\mu_{j,\sigma}|^{4\kappa-1}\theta(\mu_{j,\sigma}-\omega)\\\nonumber
G_{\psi_{j,\sigma}}^{-+}(\omega)&=&\frac{2\pi a_0^{4\kappa}}{v^{4\kappa}}\frac{-i}{\Gamma(4\kappa)}|\omega-\mu_{j,\sigma}|^{4\kappa-1}\theta(\omega-\mu_{j,\sigma})
\end{eqnarray}
with $\tilde{h}(\kappa)=2e^{-2\pi i \kappa}\sin(2\pi\kappa)\Gamma(1-2\kappa)^2$.

From Eq.(\ref{curr}) the current is related to the evaluation of $G_{\psi_{j,\sigma}}^{<}(\omega)=G_{\psi_{j,\sigma}}^{+-}(\omega)$, $G_{\psi_{j,\sigma}}^{A}(\omega)=G_{\psi_{j,\sigma}}^{++}(\omega)-G_{\psi_{j,\sigma}}^{-+}(\omega)$, the full retarded dot Green's function  
$G_{d\sigma}^{R}(\omega)=G_{d0\sigma}^{R}(\omega)+G_{d0\sigma}^{R}(\omega)\Sigma^R(\omega)G_{d\sigma}^{R}(\omega)$, and the full dot lesser Green's function
$G_{d\sigma}^{<}(\omega)=G_{d\sigma}^{R}(\omega)\Sigma^{<}(\omega)G_{d\sigma}^{A}(\omega)$. Here the bare dot retarded Green's function is
$G_{d0\sigma}^{R}(\omega)=1/(w-\epsilon_d+i0^+)$, and the dot self energy is $\Sigma(\omega)\equiv\sum_{j,\sigma}|t_{j,\sigma}|^2G_{\psi_j,\sigma}(\omega)$.

\end{document}